# Metal hydrogen sulfide superconducting temperature

Nikolay A. Kudryashov[1], Alexander A. Kutukov[1], Evgeny A. Mazur[1*)]

*Corresponding Author: E-mail: EAMazur@mephi.ru

[1] National Research Nuclear University "MEPHI", Kashirskoe sh.31, Moscow, 115409, Russia

Éliashberg theory generalized to the case of the electron-phonon (EP) systems with the not constant density of electronic states, as well as to the case of the frequency dependent renormalization of both the electron mass and the chemical potential, is used to study hydrogen sulphide $SH_3$ phase under pressure. The frequency and the temperature dependences of the electron mass renormalization $\mathrm{ReZ}(\omega)$, the density of electronic states $N(\varepsilon)$, renormalized by the EP interaction , the spectral function of the electron-phonon interaction, obtained by calculation are used to calculate the electronic abnormal GF. The generalized Éliashberg equations with the variable density of electronic states are resolved for the hydrogen sulphide $SH_3$ phase under pressure. The dependence of both the real and the imaginary part of the order parameter on the frequency in the $SH_3$ phase is obtained. The results of the solution of the Eliashberg equations for the Im-3m (170 GPa), Im-3m (200 GPa), Im-3m (225 GPa) and R3m (120 GPa) hydrogen sulphide $SH_3$ phases are presented. The $T_c = 177K$ value in the hydrogen sulfide $SH_3$ phase at the pressure P = 225 GPa has been defined. The $T_c = 241K$ value for the hydrogen sulfide $SH_3$ phase at the pressure P = 170 GPa has been predicted.

## 1. Introduction

The calculation of the superconducting transition temperature $T_c$ in the hydrogen sulphide attracted the attention of researchers due to the experiments [1-3]. Consistent consideration of this problem should include: 1. the calculation of the properties of the electronic subsystem with the full account of a non-constant electron density of states, the calculation of the properties of the phonon subsystem of the crystal and the precise calculation of the spectral function of the electron-phonon interaction. 2. Further, using these parameters the self-consistent equations describing the normal and superconducting properties of strongly interacting electron-phonon system should be solved. 3. At last it should be resolved the issue of the stability and transformation of the crystalline phases, taking into account the contribution of the zero-point fluctuations. In [4] the value of the maximum $T_c$ for the metallic $SH_2$ at 160 GPa has been estimated to be equal $80K$. In [4] the spectral function of the electron-phonon interaction for the metallic $SH_2$ has been calculated using the Espresso package. In [5] a comparative study of the enthalpy of the phases with the Im3m and R3m symmetry for the structures with $SH_3$ stoichiometry has been performed. The contribution of the zero-point fluctuations, as well as the contribution of phonons excited at a given temperature in comparing the enthalpy of competing phases in [5] has not been taken into account. The impact of the renormalization of the electron density of states on both the energy and the enthalpy of the compared phases has not also been taken into account, thus leaving the question of the dynamic stability of competing phases, for example, R3m phase, open. In [5] also a calculation of the spectral functions of the electron-phonon interaction for the $SH_3$ phases with Im3m and R3m symmetries has been carried out. In [6] the Éliashberg equations on the imaginary axis for the $H_2S$ phase have been solved, without account of the variable nature of the electron density of states, and without regard of the chemical potential renormalization. In the calculations [6] the spectral function of the electron-phonon interaction for $H_2S$ from [4] has been used. Several attempts of the solution of the Éliashberg equations with the variable density of electronic states have to date been done in the works of other authors. In the first of these works [7] not Éliashberg equations have been resolved and only equations simulating Éliashberg equation within the DFT (density functional theory). Thus the electron density functional in this approach is chosen arbitrarily. That is, there is a fit for the desired result of the calculation by trying different functionals of the electron density. In [7] the authors express the hope that the results will be available at the level of the solution of Éliashberg equations. As can be seen from Fig. 1 of the paper [7] , the authors hope has not been confirmed. The $T_c$ values found in the framework used by the authors of [7] (curve Im3m for the $SH_3$ phase) do not correspond to the results of the experiment. In the next work [8] the same research group have already appealed to the Éliashberg equations. In [8] an attempt was made to solve Éliashberg equations with the variable density of electronic states. Unfortunately, the analysis of the article [8] can not be regarded as fully consistent. Indeed, as can be seen from the equation (43) of the article [8] , the authors based their article on the ratio for the mass renormalization factor disregarding the term with the chemical potential renormalization. The authors of the article under discussion [8] have not taken into account the fact that the self-energy part of the electron Green function is expressed through the combination of two values: $\hat{\Sigma}(i\omega_m) = i\omega_m\left[1 - Z(\vec{p},\omega_m)\right]\hat{\tau}_0 + \chi(\vec{p},\omega_m)\hat{\tau}_3$, i.e. , the self-energy in [8] in reality should be expressed not only through the renormalization of the electron mass $Z(\vec{p},\omega_m)$, and in addition it should be expressed in terms of the value commonly referred to as the renormalization of the chemical potential $\chi(\vec{p},\omega_m)$. From the above it is clear that the results of the paper [8] should be revisited. In [9] the accurate calculations of the spectral function of the electron-phonon interaction for the $SH_3$ phase with Im3m symmetry at 200 and 250 GPa with the inclusion of the anharmonic effects have been performed. Unfortunately, the calculation of $T_c$ in this work was carried out using the standard version of Éliashberg equations not taking into account the variable nature of the electron density of states and the chemical potential renormalization. In [10] the normal and superconducting properties as well as the effects of the anharmonicity in the electron-phonon interaction of the phases with the Im3m and R3m symmetry for the structures with the $SH_3$ stoichiometry have been studied in a wide range of pressures. To estimate the $T_c$ value in [10] the ordinary isotropic Migdal -Eliashberg equations have been used. In [11] it is noted that conventional Éliashberg equations do not describe the superconductivity in the metal hydrogen sulfide in the proper way. In [12] the Fermiology of the Lifshitz transitions as a function of pressure in $SH_3$ has been discussed. In the work [13] the influence of the zero point motion on the electronic states in $SH_3$ superconductor has been investigated.

The calculation of the superconducting transition temperature in the hydrogen sulphide $[1-3]$ is usually carried out by using the extremely lax and insufficiently reasoned description of superconductivity with a freely selectable electron density functional $[14,15,7]$ without any possibility of taking into account the effect of the nonlinear nature of Éliashberg equations. The approximate solutions of the Éliashberg equations $[16-19]$ not taking into account the variable nature of the electronic density of states $N(\omega)$ are also used. The solution of the Éliashberg equations on the real axis is considered to be a difficult task. Therefore the critical temperature $T_c$ by the Éliashberg formalism is usually calculated in the Green's function presentation with the use of a set of discrete Green's function values on the imaginary axis and not taking into account the variable nature of the electronic density of states (see [20] for such calculation for the hydrogen sulfide).The analytic continuation of such a solution in order to determine the frequency dependence of the order parameter is extremely inaccurate. With the use of the thermodynamic Green's function formalism it is possible to calculate only the superconducting transition temperature $T_c$. Thermodynamic Green's function formalism gives no possibility to obtain the frequency dependence of the complex superconducting order parameter as well as the frequency dependence of both the electron mass renormalization and the renormalization of the chemical potential. The method of solving the Éliashberg equations on a set of discrete points of the imaginary axis, faced with the problem of the convergence of a high order discrete matrix, has not accurately reproduce the dependence of the order parameter on the frequency after the analytic continuation of the solution for the order parameter on the real frequency axis compared to the method of solving the Éliashberg equations on the real axis. Furthermore, the exact calculation of $T_c$ by means of the thermodynamic Green's function formalism faces a substantial increase in the dimension of the treated matrix, resulting in the significant computational difficulties.

The aim of present work is to derive the generalized Éliashberg equations for the precise $T_c$ calculation in the materials with the strong EP interaction, allowing a quantitative calculation and prediction of the superconducting properties and $T_c$ in the different phases of the hydrogen sulfide $[1-3]$, as well as in the high-temperature materials with the EP mechanism of superconductivity, which will be discovered in the near future. We take into account all the features of the frequency behavior of the spectral function of the EP interaction, the behavior of the electronic density of states $N_0(\omega)$ as well as the specific properties of matter in which the superconducting state is established. The derivation of the Éliashberg equations $[21-22]$ is performed anew on a more rigorous basis, leading to the new terms in the equations for the order parameter. We foresee a minor role of the Coulomb contribution to the order parameter in view of the smallness of the Coulomb pseudopotential $\mu^* \approx 0.1$ in the hydrogen sulfide $SH_3$ phase compared with the significant constant $\lambda \sim 2.273$ of the EP interaction in the sulfide phase. We construct in the present paper a revised version of the Migdal-Éliashberg theory for the EP system at nonzero $T \neq 0$ temperature in the Nambu representation with the account of several factors such as the variability of the electronic density of states $N_0(\varepsilon)$ within the band, both the frequency and the temperature dependence of the complex mass renormalization $\operatorname{Re} Z(\omega,T)$, $\operatorname{Im} Z(\omega,T)$, the frequency and the temperature dependence of the $\operatorname{Re}\chi(\omega,T)$, $\operatorname{Im}\chi(\omega,T)$ terms usually referred to as the «complex renormalization of the chemical potential», the spectral function $\alpha^2 F$ of the electron-phonon interaction obtained by calculation, as well as the effects resulting from both the electron-hole non-equivalence and the fact of the electron band width finiteness.

Under these conditions, the derivation of the Éliashberg equations $[21-22]$ is performed anew on a more rigorous basis, leading to the new terms in the equations for the order parameter, which are not accounted for in the previous versions of the theory (see., e.g. $[21-34]$). The formalism of the work $[35]$, although formally accounts for the fact of the inconstancy of the electronic density of states, however, is in our opinion too cumbersome. As it has been shown in $[36]$ in the case of the strong electron-phonon interaction the reconstruction of both the real part $\mathrm{Re}\,\Sigma$ and the imaginary part $\mathrm{Im}\,\Sigma$ of the GF self-energy part (SP) in the materials with the variable density of electronic states is not limited to the frequency $\omega_D$ domain of the limiting phonon frequency, and extends into the much larger frequency range $\omega \gg \omega_D$ from the Fermi surface. As a result, the EP interaction modifies self-energy part of the Green's function, including its anomalous part, at a considerable distance from the Fermi surface in the units of the Debye phonon frequency $\omega_D$, and not only in the vicinity $\mu-\omega_D < \omega < \mu+\omega_D$ of the Fermi surface.

**2. The theory of the normal and superconducting properties of the electron-phonon system with the not constant density of electronic states**

Given all that is written above, we consider the EP system with the Hamiltonian which includes the electronic component $\hat{H}_e$, the ionic component $\hat{H}_i$ and the component corresponding to the electron-ion interaction in the harmonic approximation $\hat{H}_{e-i}$, so that $\hat{H} = \hat{H}_e + \hat{H}_i + \hat{H}_{e-i} - \mu\hat{N}.$ Here the following notations are introduced: $\mu$ as the chemical potential, $\hat{N}$ as the operator of electron's number in the system. The matrix electron Green's function $\hat{G}$ is defined in the representation given by Nambu as follows: $\hat{G}(x,x') = -\langle T\Psi(x)\Psi^{+}(x')\rangle$ , where conventional creation and annihilation operators of electrons appear as Nambu operators. The transition from the electron-ion Hamiltonian to the electron and phonon Green's functions is performed in the present article, using the self-consistent theory of the electron-phonon systems developed by several authors $[37-40]$. We do not specify the potential of the electron-phonon interaction, and get the potential of the electron-phonon interaction within the self-consistent system for the electron and phonon Green's functions. At the same time when receiving such a self-consistent system we start from the Hamiltonian, including only the electrons and ions of the system. Writing down the standard movement equations for the electron wave functions and averaging these equations with the Hamiltonian $\hat{H}$ we will be able to obtain the system of the self consistent equations for the electron and phonon Green functions $[37-40]$ (Appendix A). The self-energy part in the matrix form $\hat{\Sigma}(x,x')$ corresponding to the EP interaction with account of the vertex function $\hat{\Gamma}$ as well as the electron-electron correlations in the normal and the superconducting state looks like:

$$\hat{\Sigma}(x,x') = -\int dx_1 \int d\boldsymbol{r}'' \frac{e^2}{|\boldsymbol{r}-\boldsymbol{r}''|} \varepsilon^{-1}(\boldsymbol{r}''\tau, x_1)\hat{\tau}_3\hat{G}(x,x_2)\hat{\tau}_3\hat{\Gamma}(x_2,x',x_1) + \sum_{n,\kappa;n',\kappa'} \int dx_3 dx_4 \times$$
$$\times \varepsilon^{-1}(x,x_3)\nabla_\alpha V_{ei\kappa}(\boldsymbol{r}_3 - \boldsymbol{R}^0_{n\kappa})\varepsilon^{-1}(x_1,x_4)\nabla_\beta V_{ei\kappa}(\boldsymbol{r}_4 - \boldsymbol{R}^0_{n'\kappa'}) D^{\alpha\beta}_{n\kappa n'\kappa'}(\tau_3-\tau_4)\hat{\tau}_3\hat{G}(x,x_2)\hat{\tau}_3\hat{\Gamma}(x_2,x',x_1). \quad (1)$$

In (1) $x \equiv \{\boldsymbol{r},t\}$ , $\boldsymbol{R}^0_{n\kappa} = \boldsymbol{R}^0_n + \boldsymbol{\rho}_\kappa$ is the radius-vector of the equilibrium position of $\kappa$ - type ion in the crystal, $V_{ei\kappa}$ is the potential of the electron-ion interaction, by the matrix symbol $\hat{\tau}_i$ ( i=0,1,2,3) the standard Pauli matrices $[28]$ are introduced. The phonon Green's function shall be defined as $D^{\alpha\beta}_{n\kappa,n'\kappa'}(\tau) = -\langle T_\tau(u^\alpha_{n\kappa}, u^\beta_{n'\kappa'})\rangle$. Here $u^\beta_{n\kappa}$ is an $\alpha$-projection of a deviation of $\kappa,n$ ion from the balance position $\mathbf{u}_{n\kappa} = \mathbf{R}_{n\kappa} - \mathbf{R}^0{}_{n\kappa}$, $\hat{\Gamma}$ is the vertex function being a matrix in $\hat{\tau}_i$ space. The vertex $\hat{\Gamma}$ behavior is supposed to be formed under the influence of the first term in (1) which incorporates all the effects of the electron-electron correlations. Hereinafter we shall not write explicitly the first term $\hat{\Sigma}_{el-el}(x,x')$ from (1), having it, however, in mind. Consider via behavior of both $\hat{\Gamma}$ and $\hat{\Sigma}_{el-el}(x,x')$ all the revealed earlier effects of the electron-electron correlations and electron-magnon interaction in the crystal. In (1) the vertex corrections to the second electron-phonon term should be neglected in accordance with the Migdal's theorem $[41]$. For the self-energy part $\hat{\Sigma}(\vec{p},i\omega_m)$ of the electron GF $\hat{g}(\vec{p},i\omega_m)$ at the frequency points $\omega_m=(2m+1)\pi T, m=0,\pm1,\pm2..$ on the imaginary axis the following decomposition on the Pauli matrixes can be written down $[28]$

$\hat{\Sigma}(\vec{p},i\omega_m)=i\omega_m\left[1-Z(\vec{p},\omega_m)\right]\hat{\tau}_0+\chi(\vec{p},\omega_m)\hat{\tau}_3+\varphi(\vec{p},\omega)\hat{\tau}_1$, so that the inverse GF should be presented in the following form $\hat{g}^{-1}(\vec{p},i\omega_m)=i\omega_m\hat{\tau}_0-\xi_{\vec{p}}\hat{\tau}_3-\hat{\Sigma}(\vec{p},i\omega_m)$. Here $\xi_{\vec{p}}$ is the electron energy not renormalized by the electron-phonon interaction. The electron momenta $\vec{p}$ are not supposed to lie on the Fermi surface. The integration is fulfilled in the entire Fermi volume and not just on the Fermi surface. Hereby with the symbol $\chi(\vec{p},\omega_m)$ the chemical potential renormalization due to the EP interaction is designated. For the normal state of the matter the value $\chi(\vec{p},\omega_m)$, which, after analytic continuation determines the frequency dependent chemical potential shift is defined by the following formula $\chi(\vec{p},\omega_m)=\frac{1}{2}\left[\Sigma(\vec{p},\omega_m)+\Sigma(\vec{p},-\omega_m)\right]$. For the normal state of the matter the value $Z(\vec{p},\omega_m)$, which, after analytic continuation determines the mass renormalization is given by the following expression

$i\omega_m\left[1-Z(\vec{p},\omega_m)\right]=\frac{1}{2}\left[\Sigma(\vec{p},\omega_m)-\Sigma(\vec{p},-\omega_m)\right]$. Replace $Z(\vec{p},\omega)$ by the quantity $Z(\xi,\omega)$

corresponding to the constant energy $\xi$. After the analytic continuation of $Z(\vec{p}, i\omega_m)$ and $\chi(\vec{p}, \omega_m)$ to the domain of the complex $\omega$ variable the functions $Z(\vec{p}, \omega)$ and $\chi(\vec{p}, \omega)$ become complex for all frequency $\omega$ values including $\omega$ values on the real axis, with the exception of the specified set of points $\omega_m = (2m+1)\pi T$. In particular, the value of $\chi(\vec{p}, \omega = 0)$ becomes complex. The electron damping on the Fermi surface is determined by the following expression $\operatorname{Im}\Sigma(\omega = 0) = \operatorname{Im}\chi(\vec{p}, \omega = 0) \neq 0.$ This term is different from zero, that determines the presence of the damping on the Fermi surface. Moreover, one can say that this damping is increasing with the increase of the temperature as with the temperature increase the value of the argument moves in the complex plane away from the nearest points $\omega = i\pi T$ and $\omega = -i\pi T$ where $\chi(\vec{p}, \omega)$ is real. With the use of the known property of the Pauli matrixes $\hat{\tau}_3\hat{\tau}_1\hat{\tau}_3 = -\hat{\tau}_1$ the following expression for the electron Green's function in the facings of the Pauli matrices is obtained [28]:

$$\hat{\tau}_3 g_R(\boldsymbol{p}, \omega_m)\hat{\tau}_3 = \frac{i\omega_m Z(\vec{p}, \omega_m)\hat{\tau}_0 + \left[\xi + \chi(\vec{p}, \omega_m)\right]\hat{\tau}_3 - \varphi(\vec{p}, \omega)\hat{\tau}_1.}{\left[i\omega_m Z(\vec{p}, \omega_m)\right]^2 + \left[\xi + \chi(\vec{p}, \omega_m)\right]^2 - \varphi^2(\vec{p}, \omega).} \tag{2}$$

Using the standard rules of the diagram technique (see, for example, [28]) we get for the electron-phonon component of the self-energy part the expression (3) written below:

$$\hat{\Sigma}^{ph}(\boldsymbol{p}, \omega_n) = -T\sum_{n'}\int \frac{d^3\boldsymbol{p}'}{(2\pi)^3}\hat{\tau}_3 g(\boldsymbol{p}', \omega_{n'})\hat{\tau}_3 \sum_j \left|g_j(\boldsymbol{p}, \boldsymbol{p}')\right|^2 D_j(\boldsymbol{p} - \boldsymbol{p}', \omega_n - \omega_{n'}). \tag{3}$$

In (3) $g_j(\boldsymbol{p}, \boldsymbol{p}') = -\sum_\kappa \frac{1}{\sqrt{2NM_\kappa \omega_j(\boldsymbol{q})}} \langle \boldsymbol{p} | \boldsymbol{e}_j(\boldsymbol{q}, \kappa) \nabla V_{ei\kappa}(\boldsymbol{r}) | \boldsymbol{p}' \rangle e^{i\boldsymbol{q}\boldsymbol{\rho}_\kappa}$, where $\omega_j(\boldsymbol{q})$ is a j-branch phonon frequency and $\boldsymbol{e}_j(\boldsymbol{q}, \kappa)$ is a polarisation vector corresponding to the momentum $\boldsymbol{q} = \boldsymbol{p} - \boldsymbol{p}'$. Use the following spectral representations for both the electron Green's function $g\left(\vec{p}, i\omega_m\right)=\int\limits_{-\infty}^{\infty} dz' a(\boldsymbol{p}, z') / \left(i\omega_m - z'\right) 2\pi$ and the phonon Green's function

$D_j(\boldsymbol{p}, \omega_n)=\int\limits_{-\infty}^{\infty} dz' b_j(\boldsymbol{p}, z') / \left(i\omega_n - z'\right) 2\pi$. Here $a(\boldsymbol{p}, z)$ is the spectral density of the thermal electron Green's function, $b_j(\boldsymbol{p}-\boldsymbol{p}', z)$ is the spectral density of a phonon Green's function. The retarded electron Green's function $g_R(\boldsymbol{p}, z)$ is related to the spectral density of the thermodynamic electron Green's function $a(\boldsymbol{p}, z)$ in the following [42] way: $a(\boldsymbol{p}, z)=2\operatorname{Im} g_R(\boldsymbol{p}, z)$. After performance of the analytical continuation $i\omega_p \rightarrow \omega + i\delta$ the self-energy part of an electron Green's function of the metal with the use of the following sum:

$$\sum_{n'} \frac{1}{i\omega_n - z'} \frac{1}{i\omega_n - i\omega_{n'} - z} = -\frac{1}{2} \frac{th\frac{z'}{2\mathrm{T}} + cth\frac{z}{2\mathrm{T}}}{i\omega_n - z - z'} \qquad (4)$$

shall be given by by the equation listed below:

$$\hat{\Sigma}^{ph}(\boldsymbol{p}, \omega)=\int \frac{d^3\boldsymbol{p}'}{(2\pi)^3} \sum_j \left|g_j(\boldsymbol{p}, \boldsymbol{p}')\right|^2 \int\limits_{-\infty}^{+\infty} \frac{dz}{2\pi} \int\limits_{-\infty}^{+\infty} \frac{dz'}{2\pi} \frac{th\frac{z'}{2\mathrm{T}} + cth\frac{z}{2\mathrm{T}}}{\omega - z - z' + i\delta} \hat{\tau}_3 \operatorname{Im} g_R(\boldsymbol{p}', z') \hat{\tau}_3 b_j(\boldsymbol{p}-\boldsymbol{p}', z). \qquad (5)$$

Integration on a pulse shall be represented as follows $\int \frac{d^3\boldsymbol{p}}{(2\pi)^3} \ldots = \int d\xi \int\limits_{S(\xi)} \frac{d^2\boldsymbol{p}}{v_{\boldsymbol{p}}} \ldots$, where $v_p$ is the electron velocity modulo at the energy surface $\xi$. Introduce a notation for the electron-phonon interaction spectral function $\alpha^2\left(\xi', \xi, z\right) F\left(\xi', \xi, z\right)$ as follows

$$\alpha^2\left(\xi', \xi, z\right) F\left(\xi', \xi, z\right)=\frac{1}{2\pi} \int\limits_{S(\xi')} \frac{d^2\boldsymbol{p}'}{v_{\mathbf{p}'}} \sum_j \left|g_j(\boldsymbol{p}, \boldsymbol{p}')\right|^2 b_j(\boldsymbol{p}-\boldsymbol{p}', z) \left( \int\limits_{S(\xi')} \Gamma(\boldsymbol{p}, \boldsymbol{p}') \frac{d^2\boldsymbol{p}'}{v_{\mathbf{p}'}} \right)^{-1}. \qquad (6)$$

The phonon contribution to the self-energy part of the electron retarded Green's function $\hat{g}_R$ with the use of the adopted notations shall be expressed in the following form:

$$\hat{\Sigma}^{ph}(\xi,\omega)=\frac{1}{2\pi}\int_{-\infty}^{+\infty}dz\int_{-\infty}^{+\infty}dz'\int_{-\mu}^{+\infty}d\xi'\alpha^2(\xi',\xi,z)F(\xi',\xi,z)N_0(\xi')\frac{th\frac{z'}{2\mathrm{T}}+cth\frac{z}{2\mathrm{T}}}{\omega-z-z'+i\delta}\hat{\tau}_3\,\mathrm{Im}\,\hat{g}_R(\xi',z')\hat{\tau}_3. \quad (7)$$

Use the technique for the real frequency Éliashberg equations. This technique allows us to control the process of calculation $\mathrm{Re}Z(\omega)$, $\mathrm{Im}Z(\omega)$, $\mathrm{Re}\Sigma(\omega)$, $\mathrm{Im}\Sigma(\omega)$, $\mathrm{Re}\chi(\omega)$ and $\mathrm{Im}\chi(\omega)$ frequency behavior. Additionally this technique gives us the opportunity to compare the results with the relevant values obtained on the experiment. In (7) $N_0(\xi)$ is a "bare" (not renormalized by the EP interaction) variable electronic density of states defined by the following expression $\int_{S(\xi)}\frac{d^2\boldsymbol{p}'}{v_{\xi\boldsymbol{p}'}}d\xi=\int_{S(\xi)}N_0(\xi)d\xi$ in which the energy of the "bare" electrons $\xi$ with the pulse $\boldsymbol{p}$ is measured from the Fermi level. It is not assumed that the electron pulses lie on the Fermi surface. Neglect in (7) the $\alpha^2F$ dependence on the $\xi$, $\xi'$ variables so that $\alpha^2(\xi',\xi,z)F(\xi',\xi,z)\approx\alpha^2(z)F(z)$. Replace the $Z(\vec{p}',\omega)$ function with $Z(\xi,\omega)$, corresponding to the constant energy $\xi$ in the direction determined by the angle $\varphi$. Average the expression (7) to the angle $\varphi$ of the pulse direction. The phonon contribution to the self energy part $\hat{\Sigma}^{ph}(\xi,\omega)$ (7) of the retarded electron GF $\hat{g}_R$ can be easily reformulated to the following form:

$$\hat{\Sigma}^{ph}(\xi,\omega)=\frac{1}{\pi}\int_{-\infty}^{+\infty}dz'\int_{-\mu}^{+\infty}d\xi'\frac{N_0(\xi')}{N_0(0)}K^{ph}(z',\omega)\hat{\tau}_3\,\mathrm{Im}\,\hat{g}_R(\xi',z')\hat{\tau}_3, \quad (8)$$

where

$$K^{ph}(z',\omega)=\int_0^{+\infty}dz\alpha^2(z)F(z)\frac{1}{2}\left\{\frac{th\frac{z'}{2\mathrm{T}}+cth\frac{z}{2\mathrm{T}}}{z'+z-\omega-i\delta}-\frac{th\frac{z'}{2\mathrm{T}}-cth\frac{z}{2\mathrm{T}}}{z'-z-\omega-i\delta}\right\}. \quad (9)$$

The $\delta$ in (9) has the infinitesimal positive value. The Coulomb contribution to the self energy part $\hat{\Sigma}(\xi,\omega)$ of the retarded electron GF $\hat{g}_R$ takes the standard [28,30-32] form:

$$\hat{\Sigma}^c(\xi,\omega)=\frac{1}{\pi}\int_{-\infty}^{+\infty}dz'th\frac{z'}{2T}\int_{-\mu}^{+\infty}d\xi'\frac{N_0(\xi')}{N_0(0)}V_c(\xi,\xi')\hat{\tau}_3\operatorname{Im}\hat{g}_R(\xi',z')\hat{\tau}_3, \qquad (10)$$

where $V_c(\xi,\xi')$ is the matrix element of the Coulomb interaction. Neglect in (8) the $\hat{\Sigma}^{ph}(\xi,\omega)$ dependence on the $\xi$ variable. Substitute $K^{ph}(z',\omega)$ (9) into (8). Further during the transition from the integration $\int_{-\infty}^{+\infty}dz'$ to the integration $\int_{0}^{+\infty}dz'$ after simple transformations we obtain formulas, describing the properties of the normal state of the crystal with a variable density of electronic states. For the normal state for the $(1,1)$ component of the matrix self-energy imaginary part $\operatorname{Im}\Sigma(\omega)=-\operatorname{Im}Z(\omega)\omega+\operatorname{Im}\chi(\omega)$ the following expression with the use of the identities $th\frac{z'}{2\mathrm{T}}=1-2f(z')=-1+2f(-z')$, $cth\frac{z}{2\mathrm{T}}=1+2n_B(z)$ may be written down:

$$\begin{aligned}&\operatorname{Im}\Sigma(\omega)=-\pi\int_0^{+\infty}dz\alpha^2(z)F(z)\times\\&\times\left\{\left[N(\omega-z)-N(\omega+z)\right]n_B(z)+N(\omega-z)f(z-\omega)+N(\omega+z)f(z+\omega)\right\}.\end{aligned} \qquad (11)$$

The expression for the real $(1,1)$ component of the matrix self-energy part $\operatorname{Re}\Sigma(\omega)=\left[1-\operatorname{Re}Z(\omega)\right]\omega+\operatorname{Re}\chi(\omega)$ has the following form:

$$\begin{aligned}&\operatorname{Re}\Sigma(\omega)=-P\int_0^{+\infty}dz\alpha^2(z)F(z)\int_0^{+\infty}dz'\left\{\left[n_B(z)+f(-z')\right]\times\right.\\&\left.\times\left(-\frac{N(-z')}{z'+z+\omega}+\frac{N(z')}{z'+z-\omega}\right)+\left[n_B(z)+f(z')\right]\left(-\frac{N(-z')}{z'-z+\omega}+\frac{N(z')}{z'-z-\omega}\right)\right\}.\end{aligned} \qquad (12)$$

In (11) , (12) $n_B(z)$ is the Bose distribution function, $f(z')$ is the Fermi distribution function. In (11), (12) $N(z')/N_0(0)$ is marked as $N(z')$ for the bravity. With the help of (11), (12) it is easy to see that $\mathrm{Re}\,\chi(\omega)$ and $\mathrm{Im}\,Z(\omega)$ are expressed in terms of the integrals, containing the difference of the density of states at the energies of the opposite sign. These terms thus reflect a measure of the electron-hole non-equivalence in the EP system, turning to zero at the zero electron-hole nonequivalence. In deriving (11), (12) the fact was taken into account that at the temperatures near $T_c$ the anomalous Green's function can be assumed to be equal zero. Given that $cth\left(\frac{\omega_{ph}}{2T}\right)\approx 1$, get as a result the formulae for the normal state, first obtained in [43]. The renormalized by the EP interaction density of the electron states $N(z')$ is expressed through the "bare" density of electron states $N_0(\xi)$ as follows:

$$N(z') = -\frac{1}{\pi}\int_{-\mu}^{\infty} d\xi' N_0(\xi')\,\mathrm{Im}\,g_R(\xi',z'). \qquad (13)$$

Such density of states as $N(z')$ is not the symmetrical (even) function of $z'$. Further during the transition from the integration $\int_{-\infty}^{+\infty} dz'$ to the integration $\int_{0}^{+\infty} dz'$ take into account the parity of both $\mathrm{Re}\,Z(z')$ and $\mathrm{Re}\,\chi(z')$ as well as the oddness of the functions $\mathrm{Im}\,Z(z')$ and $\mathrm{Im}\,\chi(z')$ simultaneously with the following order parameter property: $\varphi(-z') = \varphi^*(-z')$ [28]. For the superconducting state from (8), (10), taking into account the expression (2) for the retarded electron Green's function $\hat{g}_R(\xi',z')$ , we obtain the equations for both the real $\mathrm{Re}\,\varphi(\omega)$ and the imaginary $\mathrm{Im}\,\varphi(\omega)$ part of the anomalous Green's function self-energy part in the form of the system of two equations having the following form:

$$\operatorname{Re}\varphi(\omega)=\frac{1}{\pi}P\int_0^{+\infty}dz'\left[K^{ph}(z',\omega)-K^{ph}(-z',\omega)\right]\int_{-\mu}^{+\infty}d\xi'\frac{N_0(\xi')}{N_0(0)}\operatorname{Im}\frac{\varphi(z')}{\left[Z(z')(z')\right]^2-\varphi^2(z')-(\xi'+\chi(z'))^2}-$$
$$-\frac{\mu^*}{\pi\left(1-\mu^*\ln\frac{\omega_c}{\omega_D}\right)}\int_0^{\omega_c}dz'th\frac{z'}{2T}\int_{-\mu}^{\infty}d\xi'\frac{N_0(\xi')}{N_0(0)}\operatorname{Im}\frac{\varphi(z')}{Z^2(z')z'^2-\varphi^2(z')-(\xi'+\chi(z'))^2},\quad (14)$$

$$\operatorname{Im}\varphi(\omega)=\frac{1}{2}\int_0^{+\infty}dz'\{\ \alpha^2(|\omega-z'|)F(|\omega-z'|)\left[cth\frac{(\omega-z')}{2\mathrm{T}}+th\frac{z'}{2\mathrm{T}}\right]sign(\omega-z')-$$
$$-\alpha^2(|\omega+z'|)F(|\omega+z'|)\left[cth\frac{(\omega+z')}{2\mathrm{T}}-th\frac{z'}{2\mathrm{T}}\right]sign(\omega+z')\ \}\times \quad (15)$$
$$\times\int_{-\mu}^{+\infty}d\xi'\frac{N_0(\xi')}{N_0(0)}\operatorname{Im}\frac{\varphi(z')}{\left[Z(z')(z')\right]^2-\varphi^2(z')-(\xi'+\chi(z'))^2}$$

Here in (14) the first term describes the role of the electron-phonon interaction and the second Coulomb term appears to have the standard view $[28]$. The Coulomb pseudopotential of electrons in the hydrogen sulfide $\mu^*\approx 0.1$ is expressed through the standard approximation of the Coulomb matrix element $\mu^*=V_cN_0(0)/\left(1+V_cN_0(0)\ln\frac{E_F}{\omega_D}\right)$ $[28]$, $\omega_D\ll\omega_c\ll E_F$, $\omega_c$ is the energy efficiency range of the Coulomb interaction. In the wide-band materials such as the $SH_3$ metal hydrogen sulfide in study the equation for the real part of the order parameter $\varphi(\omega)$ for the electron-phonon systems with the variable density of electron states takes the following form (Appendix B):

$$\operatorname{Re}\varphi(\omega) = -P\int_0^{+\infty} dz'\left[K^{ph}(z',\omega) - K^{ph}(-z',\omega)\right]\frac{\operatorname{Re}\varphi(z')}{\sqrt{\operatorname{Re}^2 Z(z')z'^2 - \operatorname{Re}\varphi^2(z') + \operatorname{Im}\varphi^2(z')}}\times$$

$$\times\frac{N_0\left(-\left|\operatorname{Re}^2 Z(z')z'^2 - \operatorname{Re}\varphi^2(z') + \operatorname{Im}\varphi^2(z')\right|^{\frac{1}{2}}\right) + N_0\left(\left|\operatorname{Re}^2 Z(z')z'^2 - \operatorname{Re}\varphi^2(z') + \operatorname{Im}\varphi^2(z')\right|^{\frac{1}{2}}\right)}{2N_0(0)} -$$

$$-\frac{\mu^*}{\pi\left(1-\mu^*\ln\dfrac{\omega_c}{\omega_D}\right)}\int_0^{\omega_c} dz' th\frac{z'}{2T}\frac{\operatorname{Re}\varphi(z')}{\sqrt{\operatorname{Re}^2 Z(z')z'^2 - \operatorname{Re}\varphi^2(z') + \operatorname{Im}\varphi^2(z')}}\times$$

$$\times\frac{N_0\left(-\left|\operatorname{Re}^2 Z(z')z'^2 - \operatorname{Re}\varphi^2(z') + \operatorname{Im}\varphi^2(z')\right|^{\frac{1}{2}}\right) + N_0\left(\left|\operatorname{Re}^2 Z(z')z'^2 - \operatorname{Re}\varphi^2(z') + \operatorname{Im}\varphi^2(z')\right|^{\frac{1}{2}}\right)}{2N_0(0)}. \tag{16}$$

The equation (16) is written down with the full account of the Coulomb term (10) in the abnormal self-energy part. For the imaginary part of the order parameter $\varphi(\omega)$ following equation should be written:

$$\operatorname{Im}\varphi(\omega) = -\frac{1}{2}\int_0^{+\infty} dz'\{\ \alpha^2(|\omega-z'|)F(|\omega-z'|)\left[cth\frac{(\omega-z')}{2\mathrm{T}} + th\frac{z'}{2\mathrm{T}}\right]sign(\omega-z') -$$

$$-\alpha^2(|\omega+z'|)F(|\omega+z'|)\left[cth\frac{(\omega+z')}{2\mathrm{T}} - th\frac{z'}{2\mathrm{T}}\right]sign(\omega+z')\ \}\frac{\pi\operatorname{Re}\varphi(z')}{\sqrt{\operatorname{Re}^2 Z(z')z'^2 - \operatorname{Re}\varphi^2(z') + \operatorname{Im}\varphi^2(z')}}\times$$

$$\times\left[\frac{N_0\left(-\left|\operatorname{Re}^2 Z(z')z'^2 - \operatorname{Re}\varphi^2(z') + \operatorname{Im}\varphi^2(z')\right|^{\frac{1}{2}}\right)}{2N_0(0)} + \frac{N_0\left(\left|\operatorname{Re}^2 Z(z')z'^2 - \operatorname{Re}\varphi^2(z') + \operatorname{Im}\varphi^2(z')\right|^{\frac{1}{2}}\right)}{2N_0(0)}\right]. \tag{17}$$

Assuming the constancy of the "bare" electron density of states $N_0(\omega)$, we can pass from a system of equations (16) - (17) to the conventional system of Éliashberg equations $[21-34]$ in which the final width of the electron band, the pairing outside the Fermi-surface, the variability of the electronic density of states and the effects of electron-hole non-equivalence are all neglected. When assuming constant electronic density of states $N_0\left(\pm\left|\mathrm{Re}^2 Z(z')z'^2-\mathrm{Re}\,\varphi^2(z')+\mathrm{Im}\,\varphi^2(z')\right|^{\frac{1}{2}}\right)=const$ the equations for the complex order parameter $\varphi(\omega)$ near $T_c$ (16)-(17) are simplified to the following form:

$$\mathrm{Re}\,\varphi(\omega)=-P\int_0^{+\infty}dz'\left[K^{ph}(z',\omega)-K^{ph}(-z',\omega)\right]\frac{\mathrm{Re}\,\varphi(z')}{\sqrt{\mathrm{Re}^2 Z(z')z'^2-\mathrm{Re}\,\varphi^2(z')+\mathrm{Im}\,\varphi^2(z')}}-$$
$$-\frac{\mu^*}{\pi\left(1-\mu^*\ln\frac{\omega_c}{\omega_D}\right)}\int_0^{\omega_c}dz'th\frac{z'}{2T}\frac{\mathrm{Re}\,\varphi(z')}{\sqrt{\mathrm{Re}^2 Z(z')z'^2-\mathrm{Re}\,\varphi^2(z')+\mathrm{Im}\,\varphi^2(z')}},\qquad(18)$$

$$\mathrm{Im}\,\varphi(\omega)=-\frac{1}{2}\int_0^{+\infty}dz'\{\ \alpha^2(|\omega-z'|)F(|\omega-z'|)\left[cth\frac{(\omega-z')}{2\mathrm{T}}+th\frac{z'}{2\mathrm{T}}\right]sign(\omega-z')-$$
$$-\alpha^2(|\omega+z'|)F(|\omega+z'|)\left[cth\frac{(\omega+z')}{2\mathrm{T}}-th\frac{z'}{2\mathrm{T}}\right]sign(\omega+z')\ \}\frac{\pi\,\mathrm{Re}\,\varphi(z')}{2\sqrt{\mathrm{Re}^2 Z(z')z'^2-\mathrm{Re}\,\varphi^2(z')+\mathrm{Im}\,\varphi^2(z')}}.\qquad(19)$$

Usually already extremely oversimplified equations (18) - (19) for the superconducting order parameter are solved neglecting the imaginary part of the order parameter, that leads to the following standard [28] form:

$$\mathrm{Re}\,\varphi(\omega)=-P\int_0^{+\infty}dz'\left[K^{ph}(z',\omega)-K^{ph}(-z',\omega)\right]\frac{\mathrm{Re}\,\varphi(z')}{\sqrt{\mathrm{Re}^2 Z(z')z'^2-\mathrm{Re}\,\varphi^2(z')}}-$$
$$-\frac{\mu^*}{\pi\left(1-\mu^*\ln\frac{\omega_c}{\omega_D}\right)}\int_0^{\omega_c}dz'th\frac{z'}{2T}\frac{\mathrm{Re}\,\varphi(z')}{\sqrt{\mathrm{Re}^2 Z(z')z'^2-\mathrm{Re}\,\varphi^2(z')}}.\qquad(20)$$

From above it is clear that such a simplified form (20) of the Éliashberg equations should be used with the great caution for a quantitative description of the superconducting transition temperature $T_c$ in the specific materials.

## 3. High $T_c$ in the hydrogen sulfide as a result of the variability of the density of electronic states in the band

In the present paper we determine $T_c$ and the frequency behavior of the complex order parameter at the different temperatures by solving the Éliashberg equations in the form of a non-linear system of equations (11)-(13), (16) , (17) for the complex order parameter in the hydrogen sulfide $SH_3$ phase (Figure1) at the different pressures with the full account of the variable character of the electronic density of states.

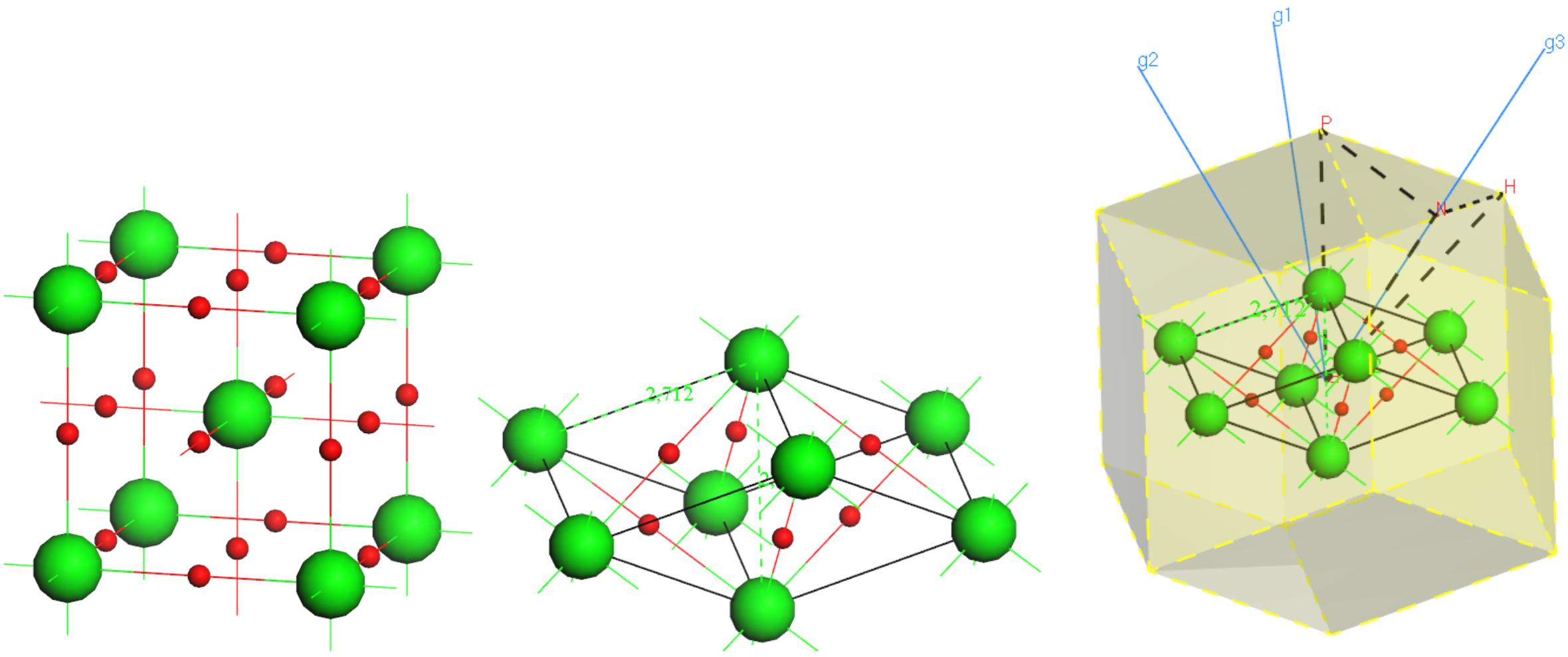

a b c

Figure1. The $SH_3$ structure investigated in the present paper [44] . Sulfur atoms are represented by a larger size balls; the results of calculations at a pressure P = 150GPa; a - the initial cubic cell; b - a primitive cell with IM-3M symmetry (OH9); c - Brillouin zone corresponding to a primitive cell.

The system of equations (11)-(13), (16) , (17) is solved by using an iterative method based on taking into account of the frequency behavior of both the bare density of $SH_3$ hydrogen sulfide electronic states (Figure 2.a) and the spectral function of the EP interaction (Éliashberg function) $\alpha^2 F(z)$ $[44,45]$ for the $SH_3$ hydrogen sulfide phase at at the different pressures (Figure 2. b).

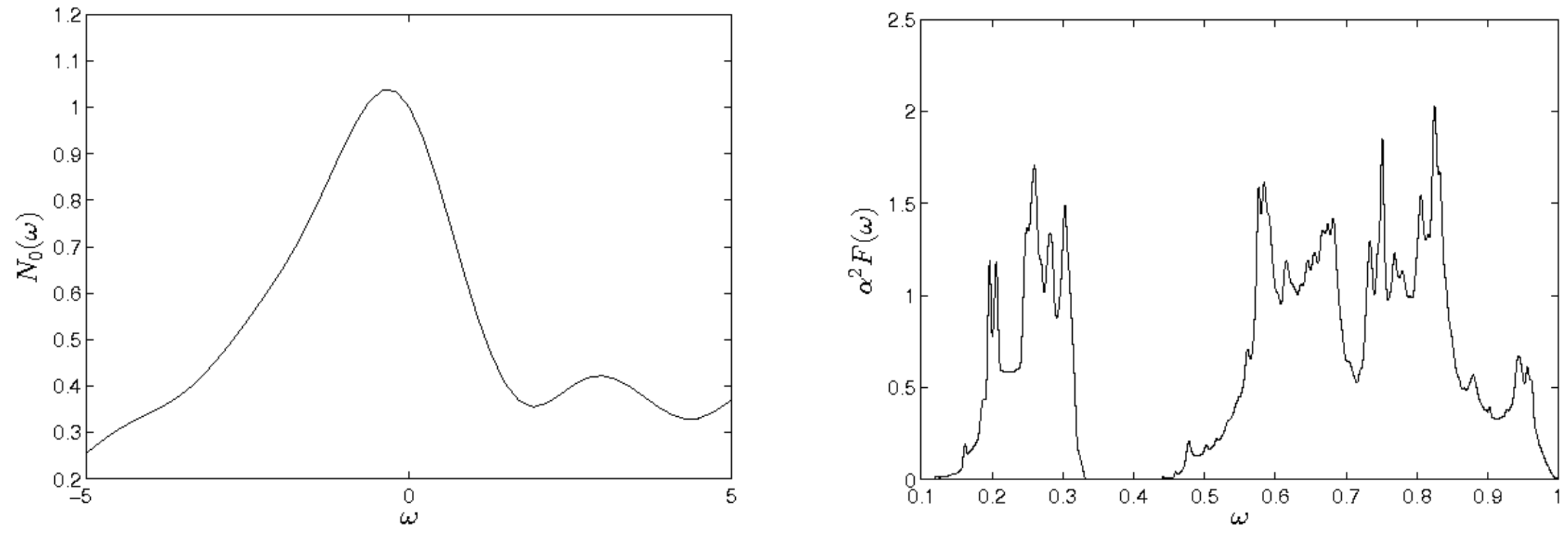

a b

Figure 2. a. Dimensionless "bare" total density of $SH_3$ hydrogen sulfide electronic states at the pressure 225 GPa [44,45]; b. The spectral function of the electron-phonon interaction $\alpha^2(\omega)F(\omega)$ [44,45] in the $SH_3$ hydrogen sulfide at the pressure 225 GPa. The frequency $\omega$ is expressed in the dimensionless units (as a fraction of the maximum frequency of the phonon spectrum).

The normal state of the $SH_3$ structure in study has been investigated in [45]. It has been found that the process of the convergence of the solution for the real part of the order parameter $\mathrm{Re}\,\varphi(\omega)$ in solving the system of equations (11)-(13), (16) , (17) is set when the number of several tens of iterations is reached. At T = 300K $\mathrm{Re}\,\varphi(\omega)$ as well as $\mathrm{Im}\,\varphi(\omega)$ tend to zero with increasing number of iterations thus indicating no effect of superconductivity at these temperature. In this case, however, the order parameter preserves the characteristic structure of the superconducting state. The order parameter is decreasing with increasing number of iterations. Equations (11)-(13), (16) ,

(17) below the $T_c$ temperature have in the case of the presence of the superconductive state a set of three solutions namely: $\mathrm{Re}\,\varphi(\omega)$ and $\mathrm{Im}\,\varphi(\omega)$, $-\mathrm{Re}\,\varphi(\omega)$ and $-\mathrm{Im}\,\varphi(\omega)$ and additionally the zero unstable solution. In the numerical solution of equations (11)-(13), (16) , (17) on the real axis the solution before the establishment of the zero solution undergoes multiple rebuilds from "negative" to the "positive" solutions. An additional difficulty in solving the equations (11)-(12), (16) , (17) is a numerical integration of improper integrals with divergences appearing in these equations. The behavior of the real part of the order parameter $\mathrm{Re}\,\varphi(\omega)$ as well as the behavior of the imaginary part of the order parameter $\mathrm{Im}\,\varphi(\omega)$ at the pressure P = 225 GPa at the temperatures T = 175K, 180K, 300K is shown in Figures 3-5. The solution for the order parameter for the resulting value of $T_c$ = 177K is not presented here due to the vanishing order parameter values at this temperature.

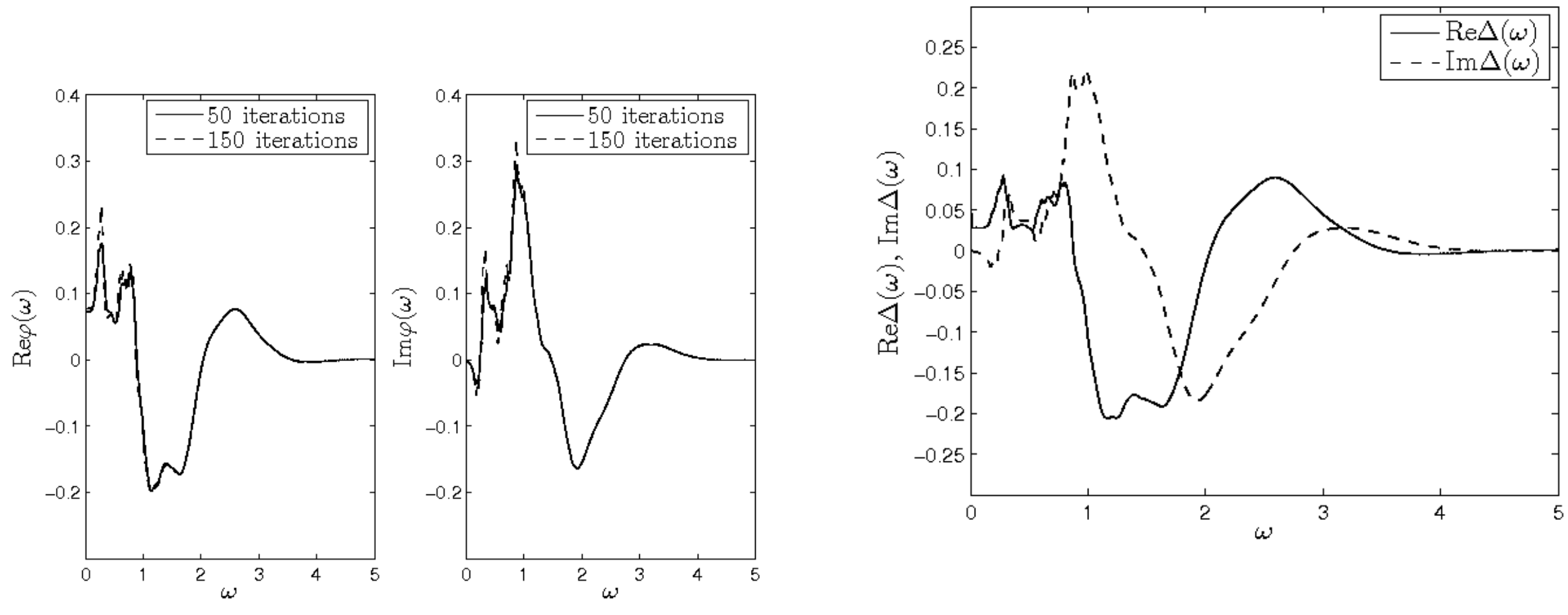

a b

Figure 3.The frequency dependence of the steady solution for a. the real part $\mathrm{Re}\,\varphi(\omega)$ and the imaginary part $\mathrm{Im}\,\varphi(\omega)$; b. the real part $\mathrm{Re}\,\Delta(\omega)$ and the imaginary part $\mathrm{Im}\,\Delta(\omega)$ of the order parameter for the $SH_3$ hydrogen sulfide phase at the temperature T = 175K and at the pressure P =

225 GPa. The frequency is expressed in the dimensionless units which correspond to the limiting frequency $0.234eV$ of the phonon spectrum.

The imaginary part $\mathrm{Im}\,\Delta(\omega)$ of the order parameter at the low frequency is negative, while at the value of the dimensionless frequency equal to 0.23 (Figure 3.b.) the imaginary part $\mathrm{Im}\,\Delta(\omega)$ becomes positive. Thus, we set the value of the energy gap in the $SH_3$ sulfide phase, which turned out to be $0.23\times 0.234eV$, that is, 600 Kelvins. Figure 4 shows the process of the $\varphi(\omega)$ vanishing with increasing number of iterations of the solution for the complex $\varphi(\omega)$ function at the temperature of T = 180K, thus indicating that $T_c < 180K$.

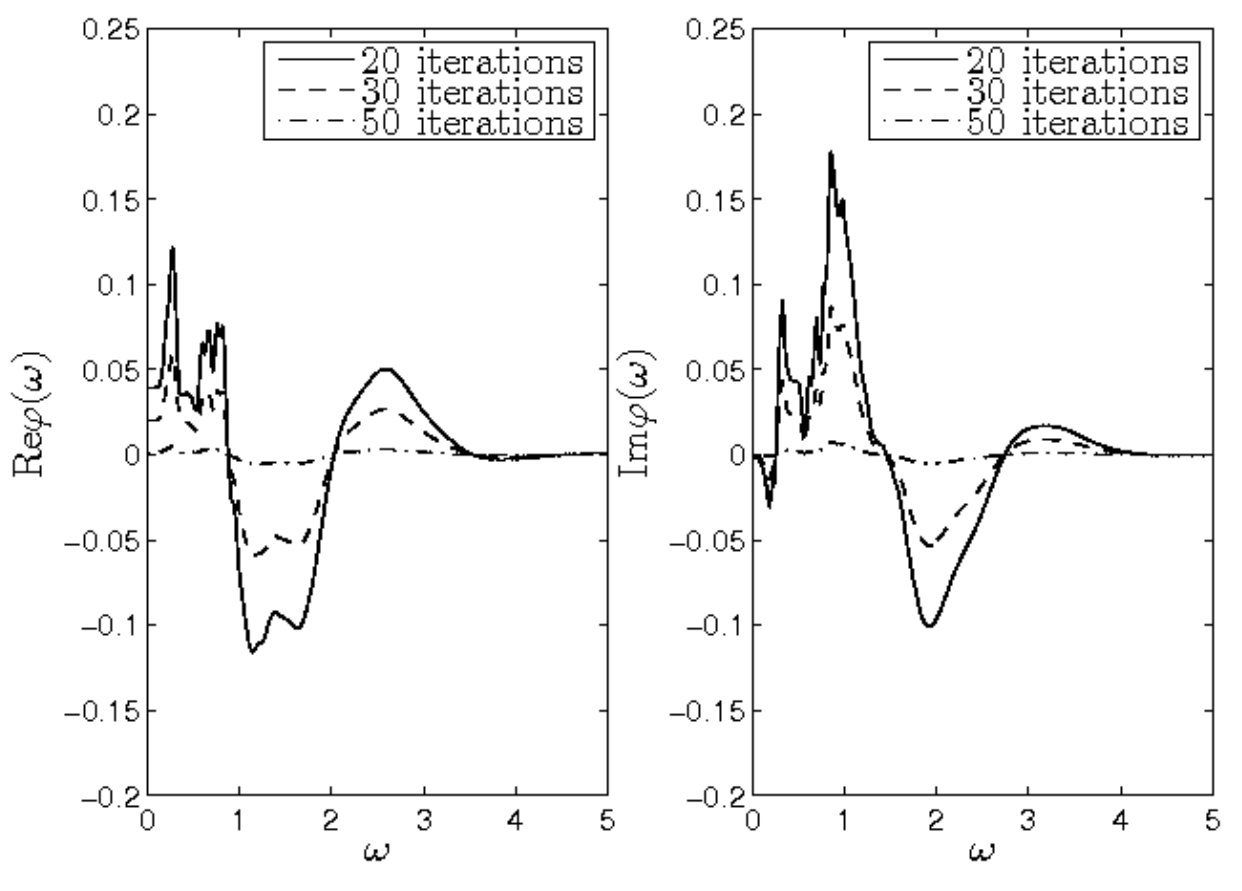

Figure 4. The dependence of the solution on the iteration number for the real part $\mathrm{Re}\,\varphi(\omega)$ and for the imaginary part $\mathrm{Im}\,\varphi(\omega)$ of the order parameter for the hydrogen sulphide $SH_3$ phase at T = 180K and at the pressure P = 225 GPa. The frequency is expressed in the dimensionless units which correspond to the limiting frequency $0.234eV$ of the phonon spectrum .

Figure 5 shows the frequency dependence at the number 100 of iterations of the very small quantities $\mathrm{Re}\,\varphi(\omega)$, $\mathrm{Im}\,\varphi(\omega)$ at the temperature T = 300K. From the Figure 5 it is clearly seen that the magnitude of the order parameter $\mathrm{Re}\,\varphi(\omega)$, $\mathrm{Im}\,\varphi(\omega)$ in the decrease with increasing number of iterations even at T = 300K retains characteristic functional dependency of the hydrogen sulfide order parameter in the superconducting state. At the same time the rough approximations of 1 for $\mathrm{Re}\,\varphi(\omega)$ and of 0 for $\mathrm{Im}\,\varphi(\omega)$ are used as the initial conditions thus introducing in the solution no functional dependency on the frequency $\omega$.

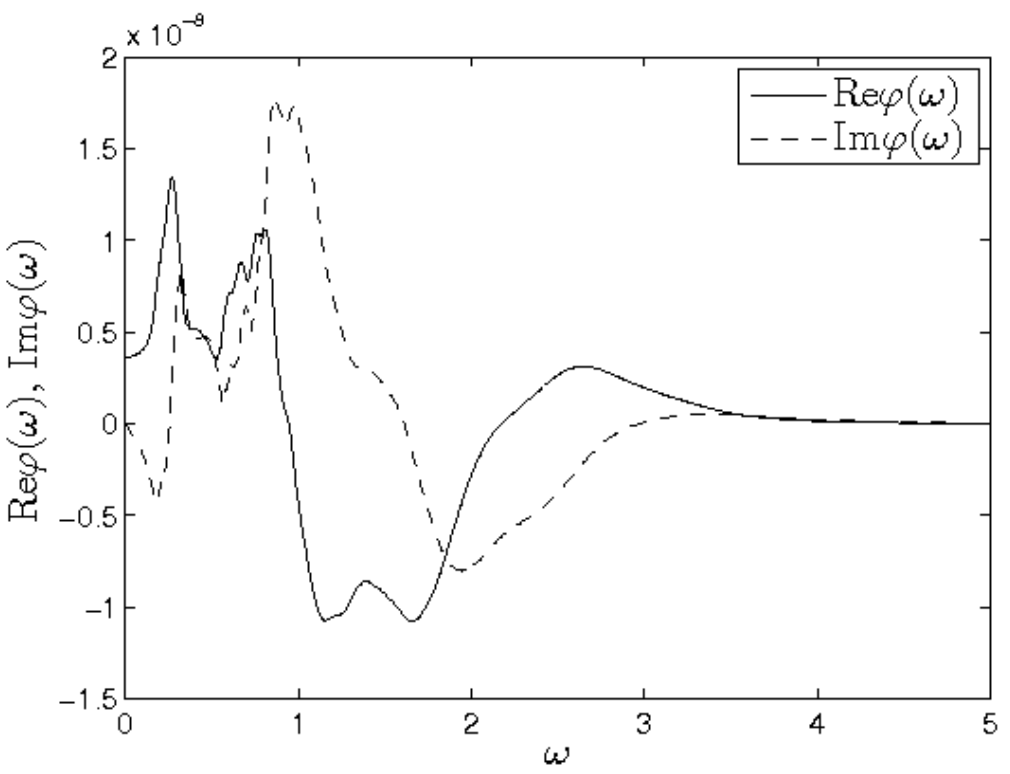

Figure 5. The frequency dependence of the solution for the order parameter at the hundredth iteration for the real part $\mathrm{Re}\,\varphi(\omega)$ and the imaginary part $\mathrm{Im}\,\varphi(\omega)$ of the hydrogen sulphide $SH_3$ phase at T = 300 K and at the pressure P = 225 GPa. The frequency is expressed in the dimensionless units, which correspond to the limiting frequency $0.234eV$ of the phonon spectrum.

Figure 6 shows the dependence of the steady-state solution for the order parameter on the temperature at the temperatures below the critical temperature $T_c$.

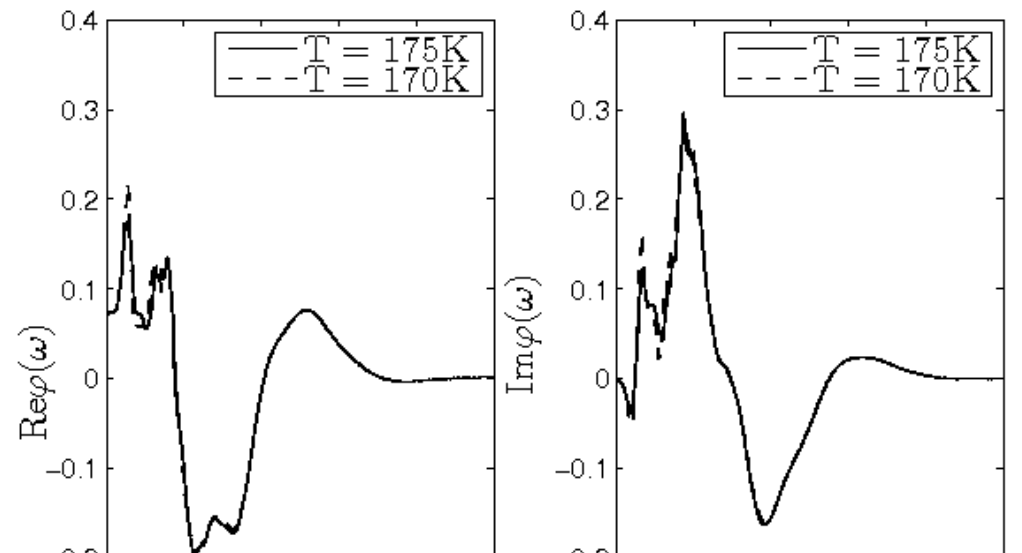

Figure 6. The temperature dependence of the steady-state solution (50 iterations) for both the order parameter real part $\mathrm{Re}\,\varphi(\omega)$ and the imaginary part $\mathrm{Im}\,\varphi(\omega)$ for the hydrogen sulfide $SH_3$ phase at the pressure of P = 225 GPa. The frequency is expressed in the dimensionless units, which correspond to the limiting frequency $0.234eV$ of the phonon spectrum.

Figure 7 shows the fundamental difference between the steady-state solutions for the complex order parameter $\varphi(\omega)$ at the temperatures below the critical temperature $T_c$ for the two possible solutions. The first of these solutions is found for the real variable relative density of electron states and the second one is obtained in the approximation of the constant density of electron states.

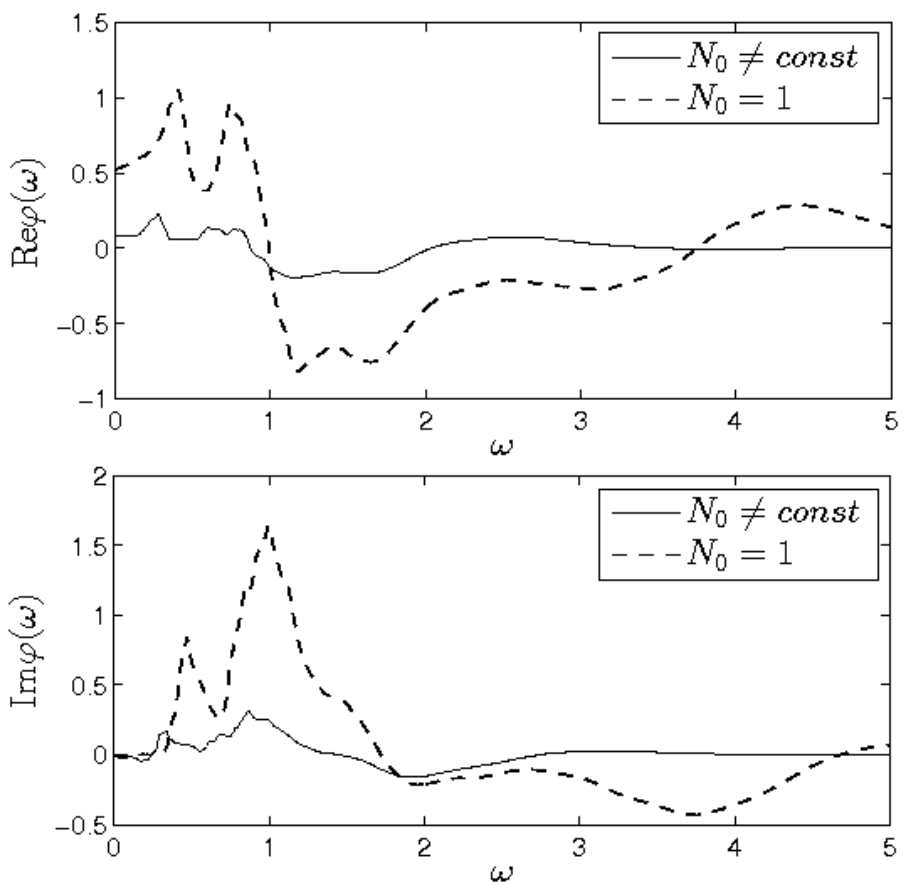

Fig. 7. The dependence of the steady-state solutions (50 iterations) on the functional form of the density of electronic states for both the order parameter real part $\mathrm{Re}\,\varphi(\omega)$ and the imaginary part $\mathrm{Im}\,\varphi(\omega)$ for the hydrogen sulfide $SH_3$ phase at the pressure of P = 225 GPa. Dotted curve: constant relative density of electron states, solid curve:real variable relative density of electron states. The frequency is expressed in the dimensionless units, which correspond to the limiting frequency $0.234 eV$ of the phonon spectrum.

One can not neglect the influence of of the chemical potential renormalization on the conduction band, because the system of equations (11)-(13), (16) , (17) includes the density of electron states renormalized selfconsistently by the electron-phonon interaction. On the Figure 8 (top part) the difference between the real part of the self energy part of the electron Green's function both with the account of the chemical potential renormalization and without account of such chemical potential renormalization is shown. The similar comparison has been made (Figure 8 (down part)) for the imaginary part of the self energy part of the electron Green's function.

Figure 8 . Top part: The real part $\mathrm{Re}\Sigma(\omega)$ of the self-energy part of the electron Green's function in the $SH_3$ hydrogen sulphide phase with the Im-3m symmetry; Down part: The imaginary part

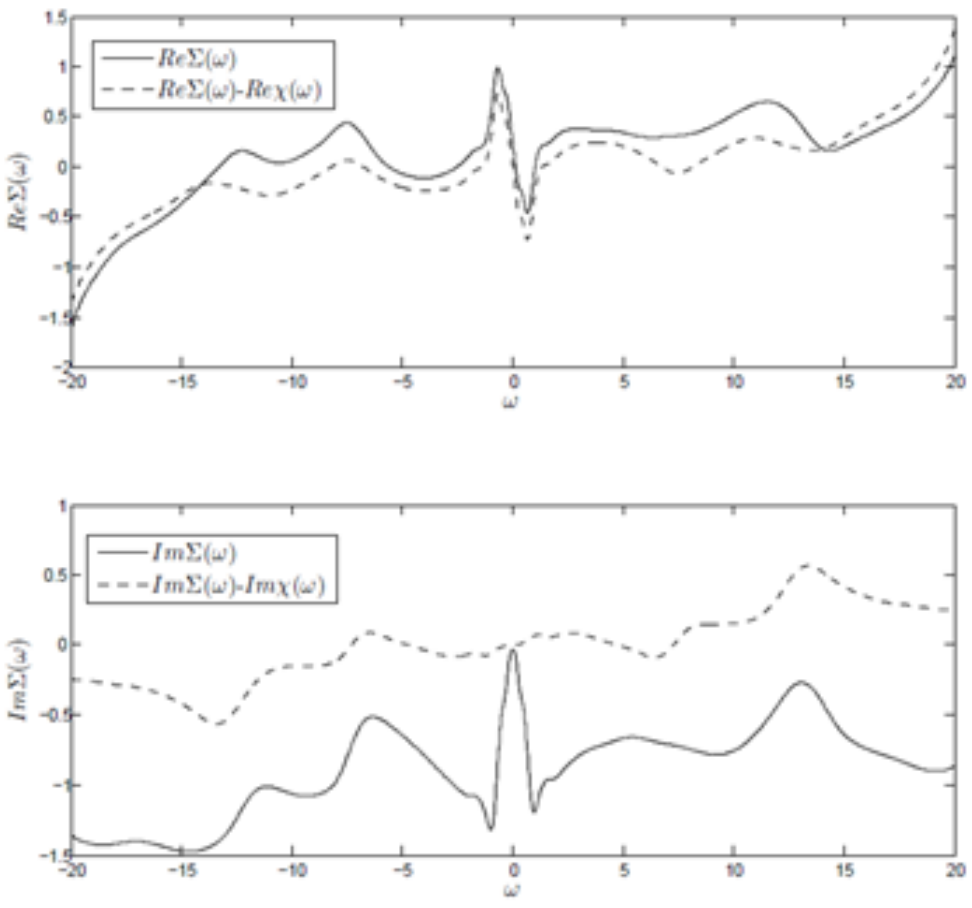

$\mathrm{Im}\,\Sigma(\omega)$ of the self-energy part of the electron Green's function. Solid line: with account of the chemical potential renormalization, dashed line: without account of the chemical potential renormalization. Both graphs are shown in the range of 4.68 electron volts at both sides of the Fermi level. Frequency $\omega$ is expressed in dimensionless units (as a fraction of the maximum frequency component 0.234 eV of the phonon spectrum for this hydrogen sulfide phase). The results are obtained for the following value of pressure P = 225 GPa (λ = 2.273) and the temperature T = 200K.

In the absence of the chemical potential renormalization the imaginary part of the self-energy part of the electron Green's function becomes odd and changes sign when passing through zero frequency. In turn, the sign of the density of electronic states is directly connected (13) with the sign of the imaginary part of the electron Green's function. Thus, the absence of the chemical potential renormalization leads to the negative values of the density of electronic states (Figure 9), which is contrary to the physical meaning. It means that both the normal and the superconducting state of a substance described by the generalized system of Éliashberg equations, should be considered only with the influence of the chemical potential renormalization.

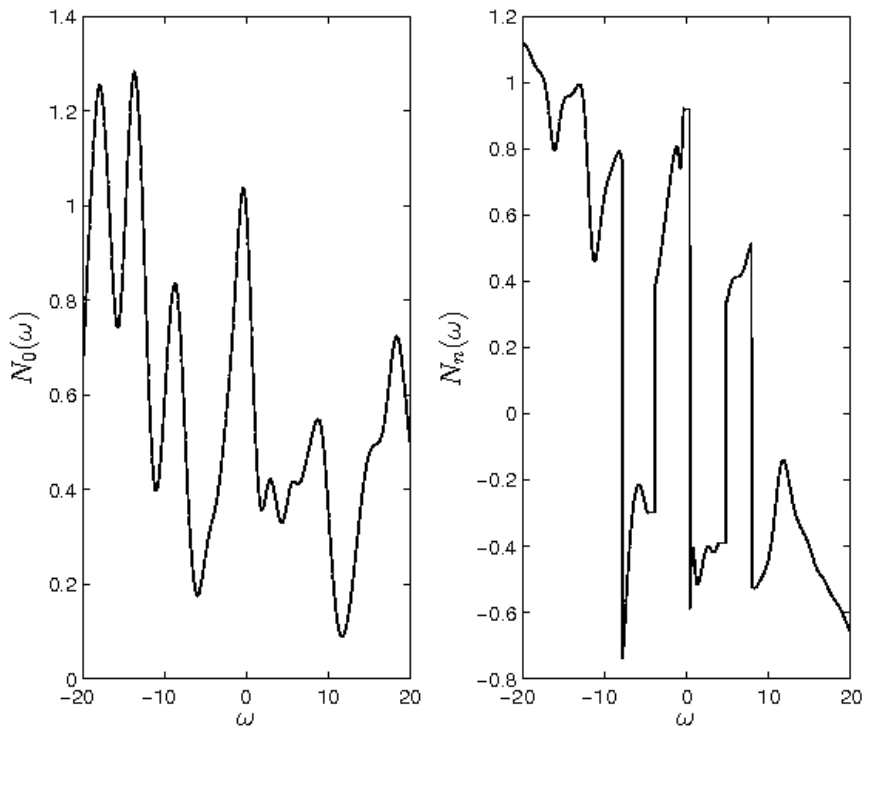

Fig. 9. Dimensionless total density of electronic states of the hydrogen sulfide $SH_3$ phase with the Im-3m symmetry. Left: "bare" electron density of states in the conduction band. To the right: reconstructed by the electron-phonon

interaction density of electronic states not taking into account the chemical potential renormalization, which correspond to the values of pressure of P = 225 GPa (λ = 2.273) and the temperature T = 200K. The frequency $\omega$ is expressed in the dimensionless units (as a fraction of the maximum frequency component 0.234 eV of the phonon spectrum for this hydrogen sulfide phase).

The results presented on the Figures 8, 9 show the drastic difference of the results with account and without account chemical potential renormalization. The solutions for both the order parameters considering the Coulomb contribution and excluding the Coulomb contribution are very little different from one another. Therefore, the graphs comparing these two solutions we do not present here. The comparison of the results for the Im-3m normal state for the three values of the pressure 170 GPa, 180 GPa and 225 GPa is presented on the Figs. 10,11:

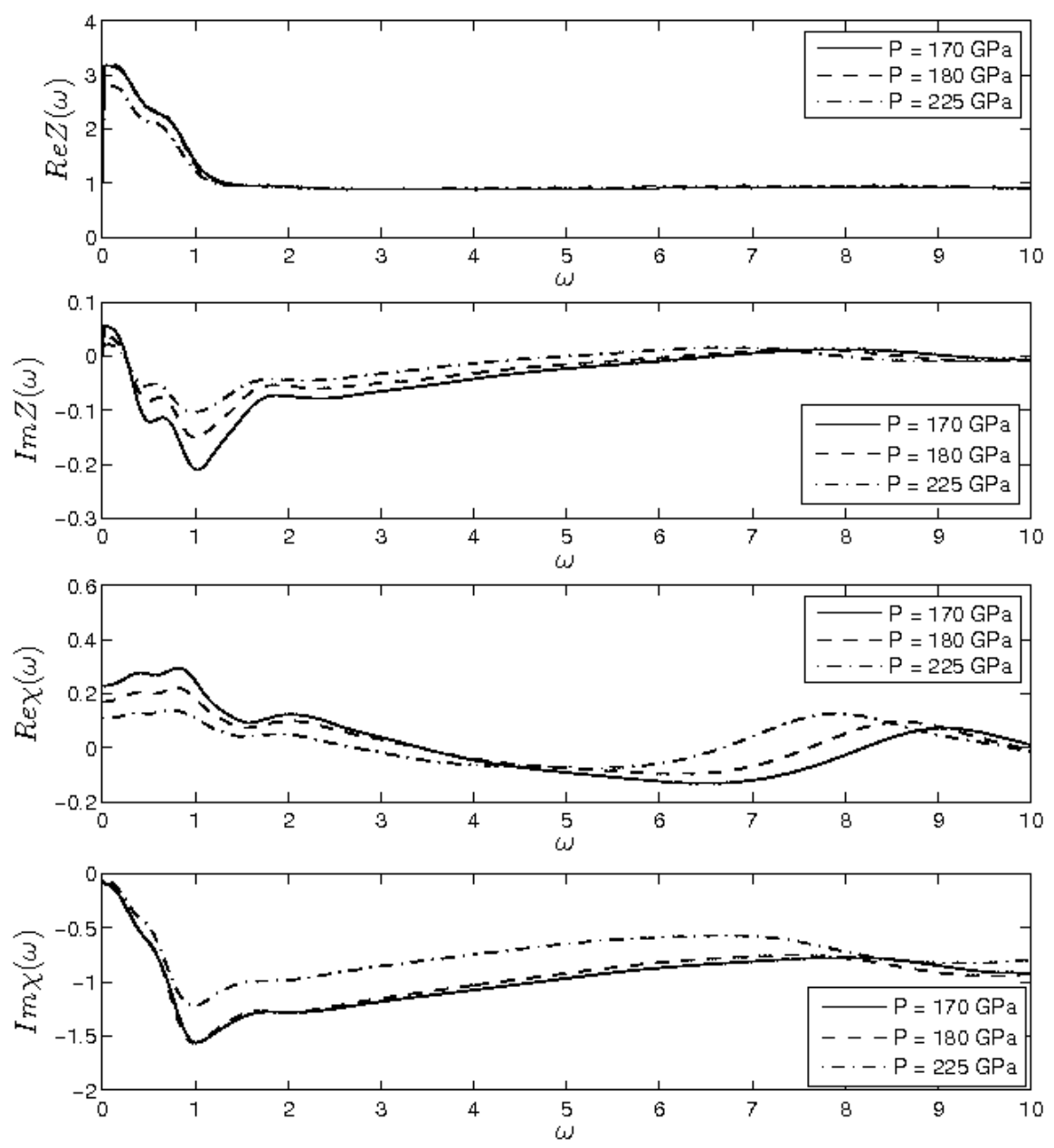

Fig. 10. Reconstructed band parameters of the conduction band of the metal $SH_3$ sulfide phase with the Im-3m symmetry. From the top to the bottom: the real part $\mathrm{Re}\,Z(\omega)$ of the electron mass renormalization $Z(\omega)$ of the Green function; the imaginary part $\mathrm{Im}\,Z(\omega)$ of the renormalization of the electron mass inside the self-energy part of the electron Green functions; renormalized with the electron-phonon interaction the real part $\mathrm{Re}\,\chi(\omega)$ of the renormalization of the chemical potential; renormalized with the electron-phonon interaction the imaginary part $\mathrm{Im}\,\chi(\omega)$ of the renormalization of the chemical potential. Frequency $\omega$ is expressed

in dimensionless units (as a fraction of the maximum frequency component 0.234 eV of the phonon spectrum for this hydrogen sulfide phase). All results are obtained for the three pressure values, namely P = 170 GPa (λ = 2.599), P = 180 GPa (λ = 2.589), P = 225 GPa (λ = 2.273) and the temperature T = 200K.

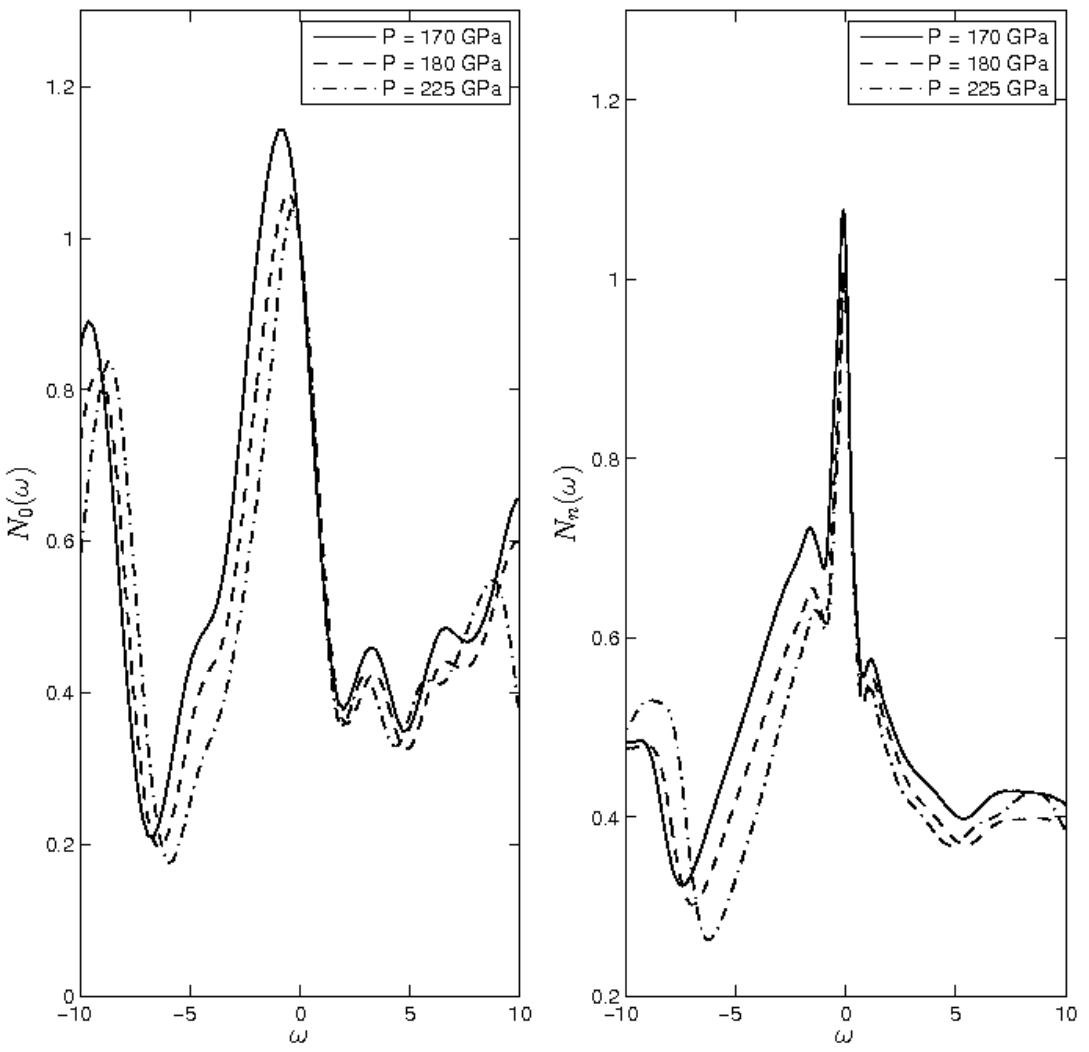

Fig. 11. Dimensionless total density of electronic states of the hydrogen sulfide $SH_3$ phase with the Im-3m symmetry. Left: "bare" electron density of states in the conduction band. To the right: reconstructed by the EP interaction density of electronic states with different values of the electron-phonon interaction constant, which correspond to the three values of pressure. All results were obtained for the three pressure values, namely: P = 170 GPa (λ = 2.599), P = 180 GPa (λ = 2.589), P = 225 GPa (λ = 2.273) and the temperature T = 200K. The frequency $\omega$ is expressed in the dimensionless units (as a fraction of the maximum frequency component 0.234 eV of the phonon spectrum for this hydrogen sulfide phase).

As can be seen from the figures 10, 11, with increasing pressure the behavior of the real $\operatorname{Re} Z(\omega)$ and the imaginary $\operatorname{Im} Z(\omega)$ parts of the complex renormalization of the electron mass, the real $\operatorname{Re}\chi(\omega)$ and imaginary $\operatorname{Im}\chi(\omega)$ parts of the complex $\chi(\omega)$ value are smoothed. The height of the renormalized peak of the electron density of states has been virtually unchanged, while the width of the peak decreases. Such changes to the characteristics of the hydrogen sulfide with pressure may be summarily described as an adverse effect of pressure on the superconductivity and may lead to a conclusion about the prospects of the research of high-temperature properties at the lower pressures. As a result of the self consistent consideration of the renormalization of the electron spectrum in the hydrogen sulphide $SH_3$ phase by the strong electron-phonon interaction, we found that the hydrogen sulfide $SH_3$ phase at a pressure of P~225 ГПа and a temperature of T = 200K is a metal with a strong non-adiabatic effects. The characteristic phonon frequency in the $SH_3$ phase is of the same order of magnitude with the characteristic width of the energy pockets of the reconstructed electronic conduction band renormalized by the strong $\lambda \approx 2.2$ electron-phonon interaction, so that $\hbar\omega_D \sim E_{cond}$ in each of these pockets.

In Figure 12 and in the Supporting information the results of the solution of the Eliashberg equations for the Im-3m (170 GPa), Im-3m (200 GPa) and R3m (120 GPa) phases are presented. For each of these parameter in addition to $T_c$ the real part $\mathrm{Re}\, Z(\omega)$ and the imaginary part $\mathrm{Im}\, Z(\omega)$ of the electron mass renormalization $Z(\omega)$, the real part $\mathrm{Re}\, \chi(\omega)$ and the imaginary part $\mathrm{Im}\, \chi(\omega)$ of the renormalization of the chemical potential are calculated. Also the "bare" electron dimensionless total density of electronic states in the conduction band together with the reconstructed by the electron-phonon interaction density of electronic states with different values of the electron-phonon interaction constant, as well as both the real part $\mathrm{Re}\, \Sigma(\omega)$ and the imaginary part $\mathrm{Im}\, \Sigma(\omega)$ of the self-energy part of the electron Green's function have been calculated. For all the specified parameters the dependence of both the real part and the imaginary part of the order parameter near $T_c$ has been calculated.

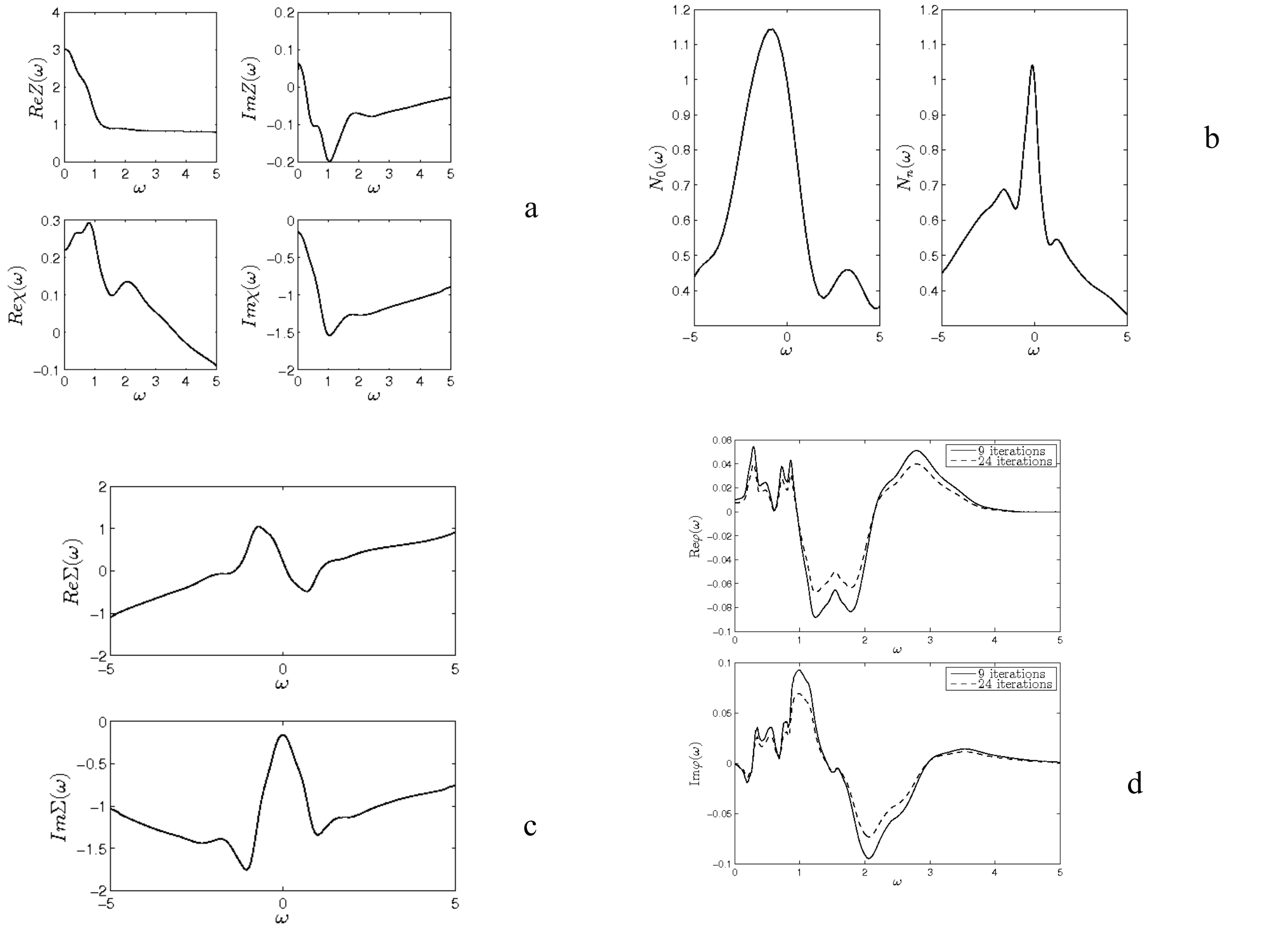

Figure 12. Reconstructed parameters of the metal $SH_3$ sulfide phase with the Im3m symmetry for the pressure P = 170 GPa (λ = 2.599) and the temperature T = 241K. a. Reconstructed band parameters of the conduction band. Top left part: the real part $\mathrm{Re}\,Z(\omega)$ of the electron mass renormalization $Z(\omega)$ of the Green function; top right part: the imaginary part $\mathrm{Im}\,Z(\omega)$ of the renormalization of the electron mass inside the self-energy part of the electron Green functions; bottom left part: renormalized by the electron-phonon interaction the real part $\mathrm{Re}\,\chi(\omega)$ of the renormalization of the chemical potential; bottom right part: renormalized by the electron-phonon interaction the imaginary part $\mathrm{Im}\,\chi(\omega)$ of the renormalization of the chemical potential; b. dimensionless total density of electronic states. Left: "bare" electron density of states in the conduction band. Right: reconstructed by the EP interaction density of electronic states with different values of the electron-phonon interaction constant. c. Top part: the real part $\mathrm{Re}\,\Sigma(\omega)$ of the self-energy part of the electron Green's function , bottom part: The imaginary part $\mathrm{Im}\,\Sigma(\omega)$ of the self-energy part of the electron Green's function. d. dependence of the solution for the real part $\mathrm{Re}\,\varphi(\omega)$ of the order parameter (top part) and for the imaginary part $\mathrm{Im}\,\varphi(\omega)$ of the order parameter (bottom part). Frequency $\omega$ is expressed in the dimensionless units (as a fraction of the maximum frequency component 0.234 eV of the phonon spectrum for this hydrogen sulfide phase).

On the Figure 13 the values of $T_c$ calculated in the present work have been imposed on the experimental plot [3] showing the dependence of the temperature of the superconducting transition $T_c$ on the pressure.

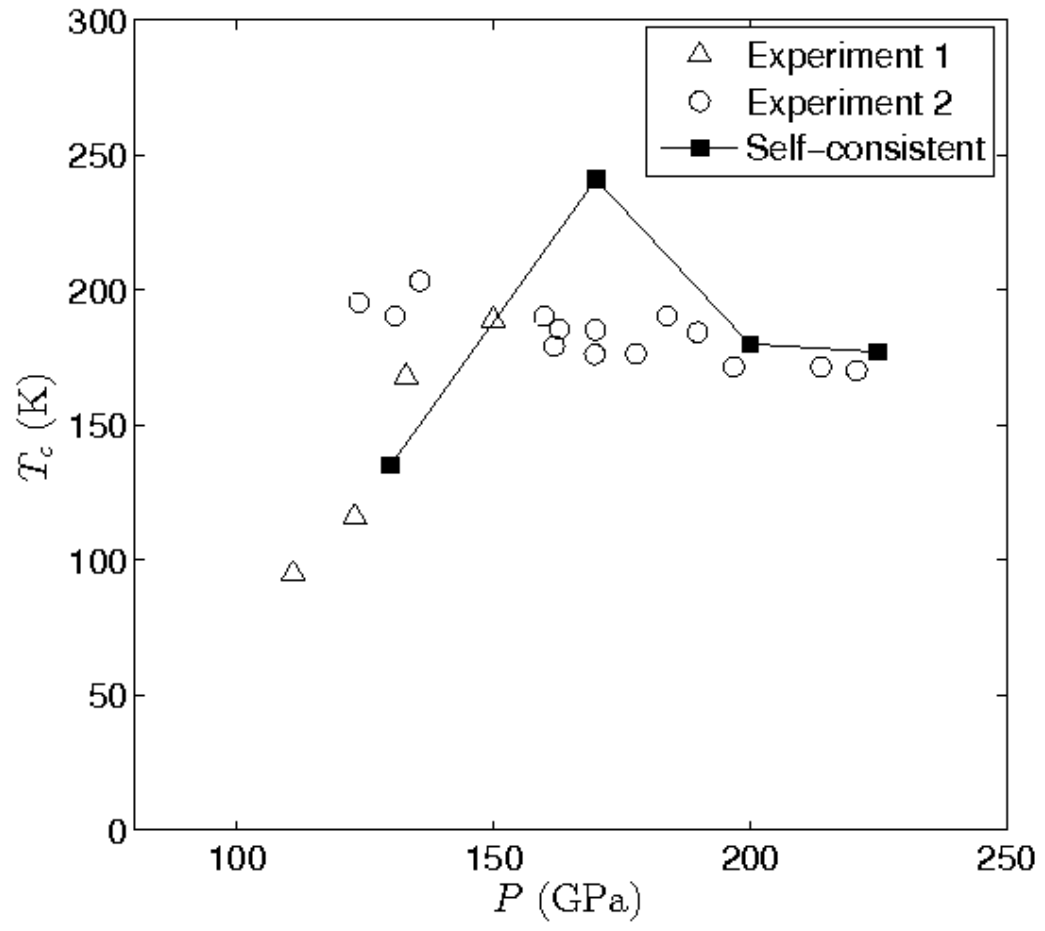

Figure 13. Dependence of $T_c$ of the $SH_3$ phase on the pressure. The results of our calculation are marked as «Self-consistent», the experimental results from the works [2,3] are denoted as «Experiment 2» and «Experiment 1», respectively.

From the Figure 13 it is seen that the calculated values for $T_c$ at the high pressures P = 180-225 GPa accurately coincide with the experimental data [2,3]. The $SH_3$ phase with the Im3m symmetry by reducing the pressure from 225 to 170 GPa, shows a steady increase in $T_c$, as is evident from the results of our calculations. The phase with the R3m symmetry also shows $T_c$ growth by increasing the pressure from 120 to 170 GPa, as is evident from the results of our calculations and the calculations [10]. At a pressure of P = 170 GPa, we predict a peak value $T_c$= 241K of the superconducting transition temperature significantly higher than the value of $T_c$, following from the experimental data [2,3]. Less pronounced tendency to increase of $T_c$ with decreasing pressure of 225 GPa to 160GPa, as well as with the pressure increase from 130 GPa to 155GPa is traced in the calculations [10 ], which take into account anharmonicity simultaneously with the use of the ordinary Eliashberg equations. Our calculations imply that at a pressure of P = 170 GPa the metal hydrogen sulfide should have the maximum temperature $T_c$ of the superconducting transition in a form of a narrow and high peak in the dependence of $T_c$ on the pressure. Probably due to the narrowness of the sharp peak or by any technological reasons the peak to date is not registered experimentally. It seems that the most accurate results for the $T_c$ in the framework of the electron-phonon mechanism of superconductivity will be further obtained with the use of an advanced version of the Éliashberg theory with both the variable nature of the density of electron states in the band and the renormalization of the electron density of states in a self-consistent form while taking into account anharmonic effects in considering phonon spectrum and the electron-phonon interaction in the framework of the approach of the present work and the work [10].

## 4. Conclusions

Analyzing the results and summing everything written before, we arrive at the following conclusions: 1. Éliashberg theory is generalized to account for the variable nature of the electron density of states in the self consistent form. 2. The mathematical method for the solving of the Éliashberg equations on the real axis is developed. 3. The generalized Éliashberg equations with the variable nature of the electron density of states are solved. The quantitative agreement with the experiment for the hydrogen sulphide $SH_3$ phase is achieved. 4. The frequency dependence as well as the fine structure of both the real part $\mathrm{Re}\,\varphi(\omega)$ and the imaginary part $\mathrm{Im}\,\varphi(\omega)$ of the order parameter corresponding to the selected $SH_3$ sulfide phase at the temperatures T = 175K , T = 180K, T = 300K has been obtained. The variation of the frequency dependence of the order parameter on the temperature is presented. 5. It is shown that at the temperatures above the critical one the order parameter is very slowly tending to zero with increasing number of iterations, maintaining the characteristic functional behavior of the superconducting state on the frequency. 6. We got $T_c = 177K$ coinsiding with the experimental $[2]$ $T_c$ value in the hydrogen sulfide at the pressure 225GPa. At the temperature $T = 180K > T_c$ and even at the room temperature T = 300 K, when the equation for the order parameter leads to the extremely small maximum values of the order parameter $\mathrm{Re}\,\varphi,\ \mathrm{Im}\,\varphi \sim 10^{-8}$ on the hundredth iteration, the order parameter frequency dependence is similar with the order parameter dependence on the frequency for the superconducting state. 6. The Coulomb pseudopotential of electrons in the hydrogen sulfide leads to the inessential $T_c$ reduction. 7. A peak value $T_c$= 241K of the superconducting transition temperature significantly higher than the experimental value of $T_c$ has been predicted.

**Supporting information. Results of the solution of Eliashberg equations for the Im-3m (200 GPa) and R3m (120 GPa) phases**

The results of the solution of Eliashberg equations for Im-3m (200 GPa) and R3m (120 GPa) phases are presented. For each of these parameter in addition to $T_c$ the real part $\operatorname{Re} Z(\omega)$ and the imaginary part $\operatorname{Im} Z(\omega)$ of the electron mass renormalization $Z(\omega)$ of the Green function, renormalized by the electron-phonon interaction,as well as the the real part $\operatorname{Re} \chi(\omega)$ and the imaginary part $\operatorname{Im} \chi(\omega)$ of the renormalization of the chemical potential are calculated. Also the "bare" electron dimensionless total density of electronic states in the conduction band together with the reconstructed by the electron-phonon interaction density of electronic states with different values of the electron-phonon interaction constant, as well as the real part $\operatorname{Re} \Sigma(\omega)$ and the imaginary part $\operatorname{Im} \Sigma(\omega)$ of the self-energy part of the electron Green's function have been calculated. For all the specified parameters the dependence of both the real part and the imaginary part of the order parameter near $T_c$ has been calculated.

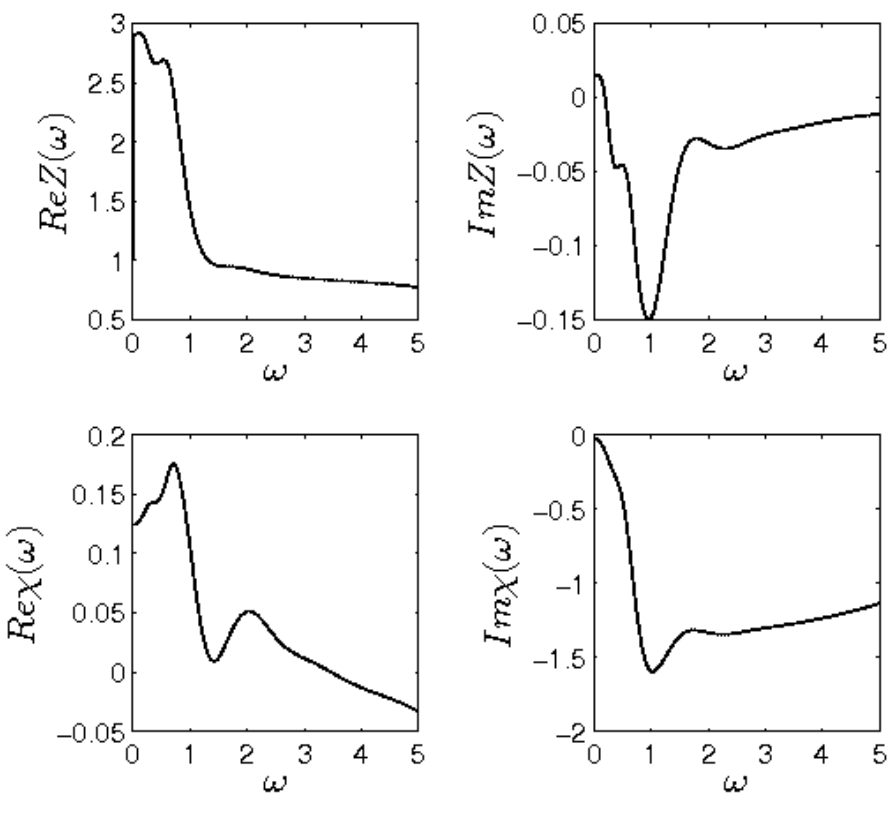

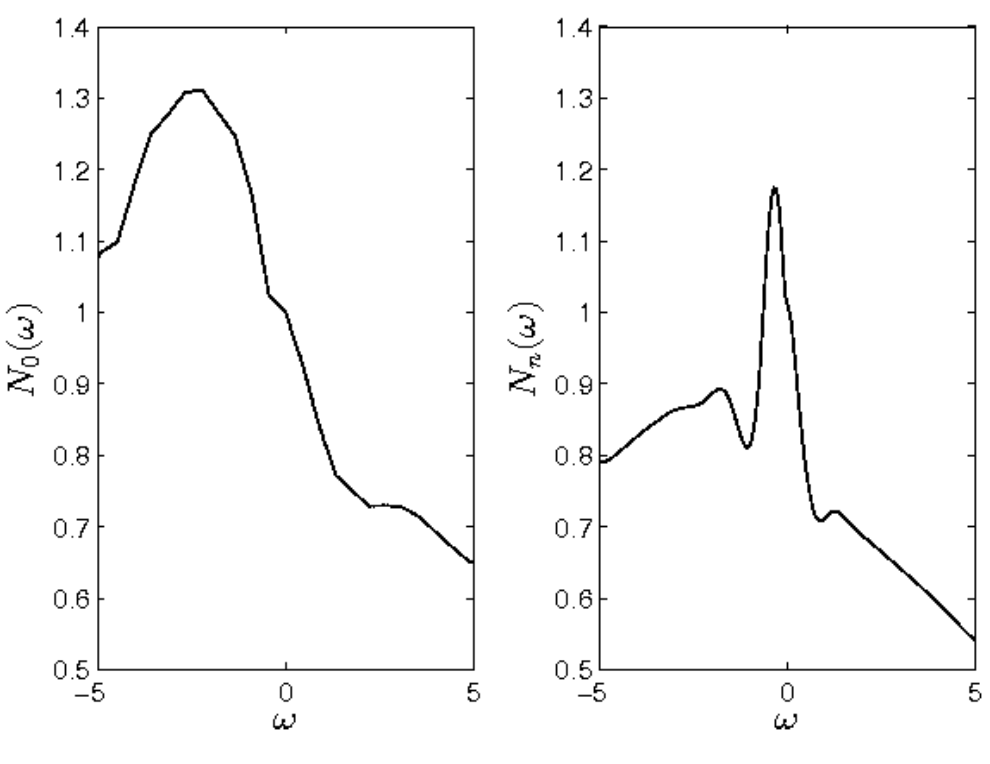

b

a

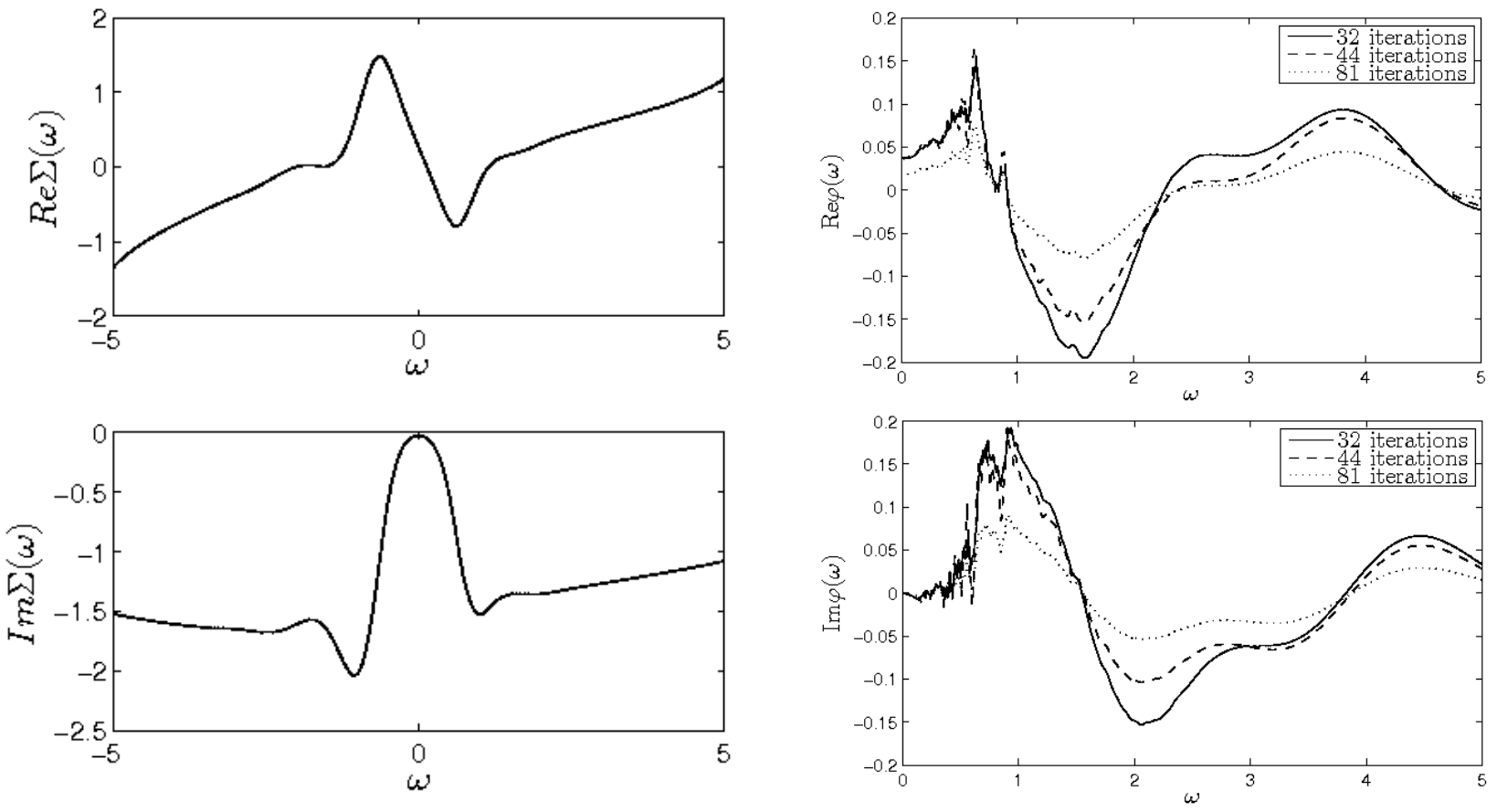

c d

Figure S1. Reconstructed parameters of the metal $SH_3$ sulfide phase with the Im3m symmetry for the pressure P = 200 GPa (λ = 2. 2.422) and the temperature T = 180K. Reconstructed band parameters of the conduction band . a. Reconstructed band parameters of the conduction band. Top left part: the real part $\mathrm{Re}\,Z(\omega)$ of the electron mass renormalization $Z(\omega)$ of the Green function; top right part: the imaginary part $\mathrm{Im}\,Z(\omega)$ of the renormalization of the electron mass inside the self-energy part of the electron Green functions; bottom left part: renormalized by the electron-phonon interaction the real part $\mathrm{Re}\,\chi(\omega)$ of the renormalization of the chemical potential; bottom right part: renormalized by the electron-phonon interaction the imaginary part $\mathrm{Im}\,\chi(\omega)$ of the renormalization of the chemical potential; b. dimensionless total density of electronic states. Left: "bare" electron density of states in the conduction band. Right: reconstructed by the EP interaction density of electronic states with different values of the electron-phonon interaction constant. c. Top part: the real part $\mathrm{Re}\,\Sigma(\omega)$ of the self-energy part of the electron Green's function ; bottom part: the imaginary part $\mathrm{Im}\,\Sigma(\omega)$ of the self-energy part of the electron Green's function. d. dependence of the solution for the real part $\mathrm{Re}\,\varphi(\omega)$ of the order parameter (top part) and for the imaginary part

$\mathrm{Im}\,\varphi(\omega)$ of the order parameter (bottom part). Frequency $\omega$ is expressed in the dimensionless units (as a fraction of the maximum frequency component 0.234 eV of the phonon spectrum for this hydrogen sulfide phase).

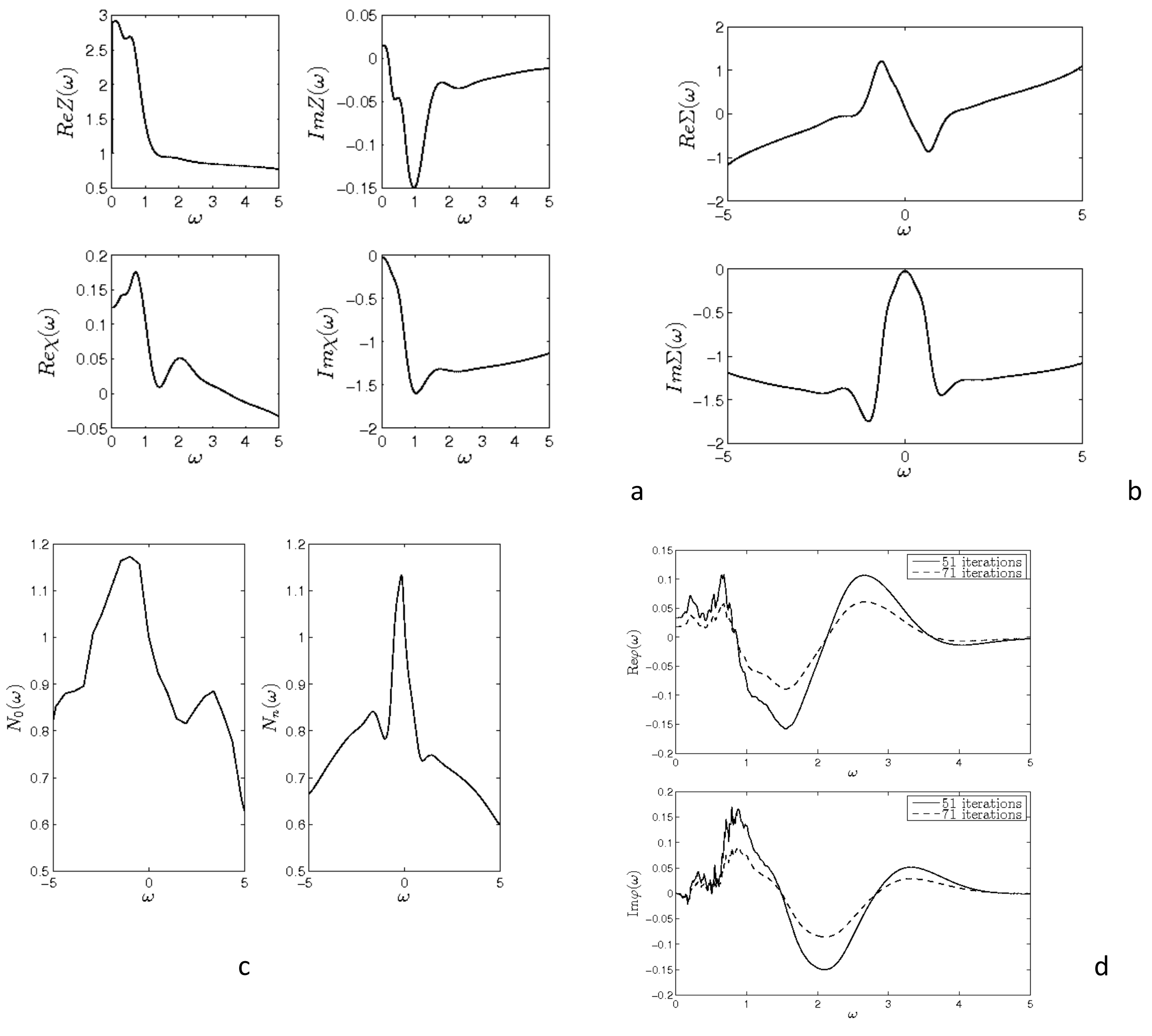

Figure S2. Reconstructed parameters of the metal $SH_3$ sulfide phase with the R3m symmetry for the pressure P = 130 GPa (λ = 2.1) and the temperature T = 135K. a. Reconstructed band parameters of the conduction band. Top left part: the real part $\mathrm{Re}\,Z(\omega)$ of the electron mass renormalization $Z(\omega)$ of the Green function; top right part: the imaginary part $\mathrm{Im}\,Z(\omega)$ of the renormalization of the electron mass inside the self-energy part of the electron Green functions; bottom left part: renormalized by the electron-phonon interaction the real part $\mathrm{Re}\,\chi(\omega)$ of the renormalization of the chemical potential; bottom right part: renormalized by the electron-phonon interaction the imaginary part $\mathrm{Im}\,\chi(\omega)$ of the renormalization of the chemical potential; b. dimensionless total density of electronic states. Left: "bare" electron density of states in the conduction band. Right: reconstructed by the EP interaction density of electronic states with different values of the electron-phonon interaction constant. c. Top part: the real part $\mathrm{Re}\,\Sigma(\omega)$ of the self-energy part of the electron Green's function , bottom part: the imaginary part $\mathrm{Im}\,\Sigma(\omega)$ of the self-energy part of the electron Green's function. d. dependence of the solution for the real part $\mathrm{Re}\,\varphi(\omega)$ of the order parameter (top part) and for the imaginary part $\mathrm{Im}\,\varphi(\omega)$ of the order parameter (bottom part). Frequency $\omega$ is expressed in the dimensionless units (as a fraction of the maximum frequency component 0.234 eV of the phonon spectrum for this hydrogen sulfide phase).

**Appendix A. Self-consistent theory of the electron-phonon systems**

In the self-consistent theory of the electron-phonon systems $[37-40]$ the expression for the electron Green's function $G(x,x')$ has the following form:

$$G^{-1}(x,x') = i\frac{\partial}{\partial\tau} + \left(\frac{\nabla^2}{2m} + \mu\right) - U_{eff} - \Sigma(x,x') \ . \qquad (A1)$$

The expression for the phonon Green's function $D^{\alpha\beta}_{nn'}(\tau)$ is written in a similar form:

$$\left(D^{\alpha\beta}_{nn'}(\tau)\right)^{-1} = \left(D^{\alpha\beta}_{0nn'}(\tau)\right)^{-1} - \Pi^{\alpha\beta}_{nn'}(\tau) . \qquad (A2)$$

The "bare" phonon Green's function $D_{0nn'}^{\alpha\beta}(\tau)$ satisfies the equation:

$$\left[M\frac{\partial^2}{\partial\tau^2}+\delta_{n''n}\sum_{n'''\neq n}\nabla_\alpha\nabla_\gamma V\left(\mathbf{R}_n^0-\mathbf{R}_{n'''}^0\right)-\sum_{n''\neq n}\nabla_\alpha\nabla_\gamma V\left(\mathbf{R}_n^0-\mathbf{R}_{n''}^0\right)-\right]D_{0n''n'}^{\alpha\beta}(\tau)=-\delta(\tau)\delta_{nn'}^{\alpha\beta}. \qquad (A3)$$

At the same time the self-energy part of the electron Green's function can be written as

$$\Sigma(x,x')=i\int dx_1\left\{\int\frac{d\vec{r}''e^2}{\left|\vec{r}-\vec{r}''\right|}\varepsilon^{-1}\left(\vec{r}''\tau,x_1\right)+\sum_{n,n'}\int dx_2dx_3dx_4\varepsilon^{-1}\left(\vec{r}\tau,x_3\right)\times\right.$$
$$\left.\times\nabla_\alpha\Phi\left(\vec{r}_3-\vec{R}_n^0\right)\varepsilon^{-1}\left(x_1,x_4\right)\nabla_\beta\Phi\left(\vec{r}_4-\vec{R}_{n'}^0\right)D_{nn'}^{\alpha\beta}\left(\tau_3-\tau_4\right)\right\}\int dx_2\Gamma\left(x_2,x',x_1\right)G\left(x,x_2\right). \qquad (A4)$$

The self-energy of the phonon Green's function $\Pi_{nn'}^{\alpha\beta}(\tau)$ is written as follows:

$$\Pi_{nn'}^{\alpha\beta}(\tau)=-\int dx''\chi_e\left(x,\vec{r}''\tau^+\right)\nabla_\alpha\Phi\left(\vec{r}-\vec{R}_n^0\right)\nabla_\beta\Phi\left(\vec{r}-\vec{R}_{n'}^0\right)+\int dx''\chi_e\left(x,x''\right)\nabla_\alpha\Phi\left(\vec{r}-\vec{R}_n^0\right)\nabla_\beta\Phi\left(\vec{r}-\vec{R}_{n'}^0\right)$$
$(A5)$

In $(4)$ the inverse matrix dielectric function $\varepsilon^{-1}(x,x')$ appears to obey the following equation:

$$\varepsilon^{-1}(x,x')=\delta(x-x')+e^2\int dx''d\vec{r}_1\frac{1}{\left|\vec{r}-\vec{r}_1\right|}\frac{\delta\left\langle\rho_e\left(\vec{r}_1,\tau\right)\right\rangle}{\delta U_{eff}(x'')}\varepsilon^{-1}(x'',x'). \qquad (A6)$$

The dielectric susceptibility $\chi_e(x,x')$ of the electrons of the crystal has the form written below:

$$\chi_e(x,x')=\int dx''\varepsilon^{-1}(x,x'')P(x'',x')=\int dx''P(x,x'')\varepsilon^{-1}(x'',x'). \qquad (A7)$$

Here $P(x,x')=\dfrac{\delta\left\langle\rho_e(x)\right\rangle}{\delta U_{eff}(x')}$, the vertex $\Gamma(x_1,x_2,x_3)$ is determined as a functional derivative

$\Gamma(x_1,x_2,x_3)=\dfrac{\delta\left\langle G^{-1}(x_1,x_2)\right\rangle}{\delta U_{eff}(x_3)}$, the dielectric susceptibility $\chi_e(x,x')$ in disregard of the retardation effects of the electronic subsystem can be supposed to depend only on the difference in the time arguments $\chi_e(x,x')=\delta(\tau-\tau')\chi_e(\mathbf{r},\mathbf{r}')$.

**Appendix B. Derivation of the Eliashberg equations**

Represent the complex functions in the following form: $Z(z')=\operatorname{Re}Z(z')+i\operatorname{Im}Z(z')$, $\chi(z')=\operatorname{Re}\chi(z')+i\operatorname{Im}\chi(z')$, $\varphi(z')=\operatorname{Re}\varphi(z')+i\operatorname{Im}\varphi(z')$. By direct calculation of the imaginary part of the complex expression, appearing in (14), (15), get the following formula:

$$\begin{aligned}
&\operatorname{Im}\frac{\varphi(z')}{\left[Z(z')(z')\right]^2-\varphi^2(z')-\left(\xi'+\chi(z')\right)^2}=\frac{1}{D}\times\\
&\times\Big\{\operatorname{Im}\varphi(z')\Big[\left(\operatorname{Re}Z^2(z')-\operatorname{Im}Z^2(z')\right)z'^2-\left(\xi'+\operatorname{Re}\chi(z')\right)^2+\left(\operatorname{Im}\chi(z')\right)^2-\operatorname{Re}\varphi^2(z')+\operatorname{Im}\varphi^2(z')\Big]-\\
&-2\operatorname{Re}\varphi(z')\left[\operatorname{Re}Z(z')\operatorname{Im}Z(z')z'^2-\operatorname{Im}\chi(z')\left(\xi'+\operatorname{Re}\chi(z')\right)-\operatorname{Re}\varphi(z')\operatorname{Im}\varphi(z')\right]\Big\}
\end{aligned}\qquad\text{(B1)}$$

Here

$$\begin{aligned}
D=&\Big[\left(\operatorname{Re}Z^2(z')-\operatorname{Im}Z^2(z')\right)z'^2-\left(\xi'+\operatorname{Re}\chi(z')\right)^2+\left(\operatorname{Im}\chi(z')\right)^2-\operatorname{Re}\varphi^2(z')+\operatorname{Im}\varphi^2(z')\Big]^2+\\
&+4\left[\operatorname{Re}Z(z')\operatorname{Im}Z(z')z'^2-\operatorname{Im}\chi(z')\left(\xi'+\operatorname{Re}\chi(z')\right)-\operatorname{Re}\varphi(z')\operatorname{Im}\varphi(z')\right]^2.
\end{aligned}\qquad\text{(B2)}$$

Neglecting both the $\operatorname{Re}\Sigma(\omega)$ and $\operatorname{Im}\Sigma(\omega)$ dependence on $\xi$ and omitting the small (as will be seen later) terms $\operatorname{Im}Z(\omega)$, $\operatorname{Re}\chi(\omega)$, $\operatorname{Im}\chi(\omega)$ in (B1), (B2) , we obtain the following non-linear expression for the real part of the complex phonon component of the abnormal self-energy part $\operatorname{Re}\sum_{abn}^{ph}(\omega)$ of the GF:

$$\begin{aligned}
&\operatorname{Re}\sum\nolimits_{abn}^{ph}(\omega)=-\frac{1}{\pi}P\int_0^{+\infty}dz'\left[K^{ph}(z',\omega)-K^{ph}(-z',\omega)\right]\times\\
&\times\int_{-\mu}^{+\infty}d\xi'\frac{N_0(\xi')}{N_0(0)}\frac{2\operatorname{Re}\varphi^2(z')\operatorname{Im}\varphi(z')}{\left\{\left[\operatorname{Re}Z(z')\right]^2z'^2-(\xi')^2-\operatorname{Re}\varphi^2(z')+\operatorname{Im}\varphi^2(z')\right\}^2+4\left[\operatorname{Re}\varphi(z')\operatorname{Im}\varphi(z')\right]^2}-
\end{aligned}$$

$$-\frac{1}{\pi}P\int_{0}^{+\infty}dz'\left[K^{ph}(z',\omega)-K^{ph}(-z',\omega)\right]\times$$

$$\times\int_{-\mu}^{+\infty}d\xi'\frac{N_0(\xi')}{N_0(0)}\frac{\operatorname{Im}\varphi(z')\left\{\left[\operatorname{Re}Z(z')\right]^2z'^2-(\xi')^2-\operatorname{Re}\varphi^2(z')+\operatorname{Im}\varphi^2(z')\right\}}{\left\{\left[\operatorname{Re}Z(z')\right]^2z'^2-(\xi')^2-\operatorname{Re}\varphi^2(z')+\operatorname{Im}\varphi^2(z')\right\}^2+4\left[\operatorname{Re}\varphi(z')\operatorname{Im}\varphi(z')\right]^2}. \quad \text{(B3)}$$

For the imaginary part of the complex phonon component of the abnormal self-energy part $\operatorname{Im}\sum_{abn}^{ph}(\omega)$ of the GF we similarly obtain the following expression:

$$\operatorname{Im}\sum\nolimits_{abn}^{ph}(\omega)=-\frac{1}{2}\int_{0}^{+\infty}dz'\{\ \alpha^2(|\omega-z'|)F(|\omega-z'|\omega-z')\left[cth\frac{(\omega-z')}{2\mathrm{T}}+th\frac{z'}{2\mathrm{T}}\right]sign(\omega-z')-$$

$$-\alpha^2(|\omega+z'|)F(|\omega+z'|)\left[cth\frac{(\omega+z')}{2\mathrm{T}}-th\frac{z'}{2\mathrm{T}}\right]sign(\omega+z')\ \}\times$$

$$\times\{\ \int_{-\mu}^{+\infty}d\xi'\frac{N_0(\xi')}{N_0(0)}\frac{2\operatorname{Re}\varphi^2(z')\operatorname{Im}\varphi(z')}{\left\{\left[\operatorname{Re}Z(z')\right]^2z'^2-(\xi')^2-\operatorname{Re}\varphi^2(z')+\operatorname{Im}\varphi^2(z')\right\}^2+4\left[\operatorname{Re}\varphi(z')\operatorname{Im}\varphi(z')\right]^2}+$$

$$+\int_{-\mu}^{+\infty}d\xi'\frac{N_0(\xi')}{N_0(0)}\frac{\operatorname{Im}\varphi(z')\left\{\left[\operatorname{Re}Z(z')\right]^2z'^2-(\xi')^2-\operatorname{Re}\varphi^2(z')+\operatorname{Im}\varphi^2(z')\right\}}{\left\{\left[\operatorname{Re}Z(z')\right]^2z'^2-(\xi')^2-\operatorname{Re}\varphi^2(z')+\operatorname{Im}\varphi^2(z')\right\}^2+4\left[\operatorname{Re}\varphi(z')\operatorname{Im}\varphi(z')\right]^2}\ \}. \quad \text{(B4)}$$

Near $T_c$ the term $\operatorname{Re}\varphi(z')\operatorname{Im}\varphi(z')$ tends to zero for all values of the argument $z'$, so that the expression

$$\frac{2\operatorname{Re}\varphi(z')\operatorname{Im}\varphi(z')}{\left\{\left[\operatorname{Re}Z(\xi',z')\right]^2z'^2-(\xi')^2-\operatorname{Re}\varphi^2(z')+\operatorname{Im}\varphi^2(z')\right\}^2+4\left[\operatorname{Re}\varphi(z')\operatorname{Im}\varphi(z')\right]^2}$$ becomes delta

function $\pi\delta\left\{\left[\operatorname{Re}Z(z')^2z'^2-(\xi')^2-\operatorname{Re}\varphi^2(z')+\operatorname{Im}\varphi^2(z')\right]^2\right\}\operatorname{sgn}\left[\operatorname{Re}\varphi(z')\operatorname{Im}\varphi(z')\right]$, whereas the

expression $\frac{\left\{\left[\operatorname{Re}Z(\xi',z')\right]^2z'^2-(\xi')^2-\operatorname{Re}\varphi^2(z')+\operatorname{Im}\varphi^2(z')\right\}}{\left\{\left[\operatorname{Re}Z(\xi',z')\right]^2z'^2-(\xi')^2-\operatorname{Re}\varphi^2(z')+\operatorname{Im}\varphi^2(z')\right\}^2+4\left[\operatorname{Re}\varphi(z')\operatorname{Im}\varphi(z')\right]^2}$

in the second term of (B3), (B4) simplifies to the following form:

$\left\{\operatorname{Re}Z(z')^2z'^2-\operatorname{Re}\varphi^2(z')+\operatorname{Im}\varphi^2(z')-\xi'^2\right\}^{-1}$. As a result the nonlinear expression (B4) for the real part of the complex phonon component of the abnormal self-energy part $\mathrm{Re}\sum_{abn}^{ph}(\omega)$ of the GF after integration with respect to $\xi'$ with the account of the delta- function properties is taken in the following form:

$$\mathrm{Re}\sum_{abn\ \mathrm{ph}}(\omega)=-P\int_{0}^{+\infty}dz'\left[K^{ph}(z',\omega)-K^{ph}(-z',\omega)\right]\frac{\mathrm{Re}\varphi(z')}{\sqrt{\mathrm{Re}^2 Z(z')z'^2-\mathrm{Re}\varphi^2(z')+\mathrm{Im}\varphi^2(z')}}\times$$
$$\times\frac{N_0\left(-\left|\mathrm{Re}^2 Z(z')z'^2-\mathrm{Re}\varphi^2(z')+\mathrm{Im}\varphi^2(z')\right|^{\frac{1}{2}}\right)+N_0\left(\left|\mathrm{Re}^2 Z(z')z'^2-\mathrm{Re}\varphi^2(z')+\mathrm{Im}\varphi^2(z')\right|^{\frac{1}{2}}\right)}{2N_0(0)}- \quad \text{(B5)}$$
$$-\frac{1}{\pi}P\int_{0}^{+\infty}dz'\left[K^{ph}(z',\omega)-K^{ph}(-z',\omega)\right]\int_{-\mu}^{+\infty}d\xi'\frac{N_0(\xi')}{N_0(0)}\frac{\mathrm{Im}\varphi(z')}{\mathrm{Re}Z(z')^2 z'^2-\mathrm{Re}\varphi^2(z')+\mathrm{Im}\varphi^2(z')-\xi'^2}.$$

For the imaginary part of the complex phonon component of the abnormal self-energy part $\mathrm{Im}\sum_{abn}^{ph}(\omega)$ of the GF we obtain as a result of the integration with respect to $\xi$ in (B4) with the account of the delta-function properties the following expression:

$$\mathrm{Im}\sum_{abn}^{ph}(\omega)=-\frac{1}{2}\int_{0}^{+\infty}dz'\{\ \alpha^2(|\omega-z'|)F(|\omega-z'|)\left[cth\frac{(\omega-z')}{2\mathrm{T}}+th\frac{z'}{2\mathrm{T}}\right]sign(\omega-z')-$$
$$-\alpha^2(|\omega+z'|)F(|\omega+z'|)\left[cth\frac{(\omega+z')}{2\mathrm{T}}-th\frac{z'}{2\mathrm{T}}\right]sign(\omega+z')\ \}\left\{-\frac{\pi\,\mathrm{Re}\varphi(z')}{\sqrt{\mathrm{Re}^2 Z(z')z'^2-\mathrm{Re}\varphi^2(z')+\mathrm{Im}\varphi^2(z')}}\times\right.$$
$$\times\left[\frac{N_0\left(-\left|\mathrm{Re}^2 Z(z')z'^2-\mathrm{Re}\varphi^2(z')+\mathrm{Im}\varphi^2(z')\right|^{\frac{1}{2}}\right)}{2N_0(0)}+\frac{N_0\left(\left|\mathrm{Re}^2 Z(z')z'^2-\mathrm{Re}\varphi^2(z')+\mathrm{Im}\varphi^2(z')\right|^{\frac{1}{2}}\right)}{2N_0(0)}\right]+ \quad \text{(B6)}$$
$$\left.-P\int_{-\mu}^{+\infty}d\xi'\frac{N_0(\xi')}{N_0(0)}\frac{\mathrm{Im}\varphi(z')}{\left[\mathrm{Re}Z(\xi',z')\right]^2 z'^2-(\xi')^2-\mathrm{Re}\varphi^2(z')+\mathrm{Im}\varphi^2(z')}\ \right\}.$$

The right hand side of equations (B5) - (B6) for the complex order parameter $\varphi(\omega)$ contains previously $[21-34]$ not taken into account contributions proportional to $\mathrm{Im}\varphi(z')$. The order parameter shall be written in the following form $\varphi(\omega)=\Delta(\omega)|Z(\omega)|$,

$\left|Z(z')\right| = \left(\mathrm{Re}^2 Z(z') + \mathrm{Im}^2 Z(z')\right)^{\frac{1}{2}}$. The $z'$ integral in (B4), (B5) is taken as a principal value, that is marked with the P mark, the negative $-\left|\mathrm{Re}^2 Z(z') z'^2 - \mathrm{Re}\,\varphi^2(z') + \mathrm{Im}\,\varphi^2(z')\right|^{\frac{1}{2}}$ value can not be less than $-\mu$, so that the integration over $z'$ at the negative $z'$ ends provided that $\left|\mathrm{Re}^2 Z(z') z'^2 - \mathrm{Re}\,\varphi^2(z') + \mathrm{Im}\,\varphi^2(z')\right|^{\frac{1}{2}} = \mu$. The integrand is equal to zero under such $z'$ that $\mathrm{Re}^2 Z(z') z'^2 - \mathrm{Re}\,\varphi^2(z') + \mathrm{Im}\,\varphi^2(z') < 0$. The root is assumed to be positive $\sqrt{\mathrm{Re}^2 Z(z') z'^2 - \mathrm{Re}\,\varphi^2(z') + \mathrm{Im}\,\varphi^2(z')} \geq 0$ for any $z'$ sign. In the wide-band materials such as the $SH_3$ metal hydrogen sulfide in study the logarithmic term in the last two equations with the high accuracy can be set equal to zero.

**References**

[1] A.P. Drozdov, M. I. Eremets, I. A. Troyan, arXiv:1412.0460, Conventional superconductivity at 190 K at high pressures.

[2] A.P. Drozdov, M.I. Eremets, I.A.Troyan, V. Ksenofontov, S.I. Shylin, Conventional superconductivity at 203K at high pressures, Nature (2015) 525, 73-76.

[3] Mari Einaga, Masafumi Sakata, Takahiro Ishikawa, Katsuya Shimizu, Mikhail Eremets, Alexander Drozdov, Ivan Troyan, Naohisa Hirao, Yasuo Ohishi, Crystal Structure of 200 K-Superconducting Phase of Sulfur Hydride System, arXiv:1509.031.

[4] Y. Li, J. Hao, H. Liu, Y. Li, and Y. Ma, J. Chem. Phys. 140, 174712 (2014).

[5] D. Duan, Y. Liu, F. Tian, D. Li, X. Huang, Z. Zhao, H. Yu, B. Liu, W. Tian, and T. Cui, Scientific Reports, **4**, 6968 (2014).

[6] A. P. Durajski, Eur. Phys. J. **B87**, 211 (2014).

[7] Ryosuke Akashi, Mitsuaki Kawamura, Shinji Tsuneyuki, Yusuke Nomura, RyotaroArita, Fully non-empirical study on superconductivity in compressed sulfur hydrides. Phys. Rev. **B 91**, 224513 (2015).

[8] Wataru Sano, Takashi Koretsune, Terumasa Tadano, Ryosuke Akashi, RyotaroArita. Effect of van Hove singularities on high-Tc superconductivity in H3S. Phys. Rev. **B 93**, 094525 (2016).

[9] I. Errea, M. Calandra, C. J. Pickard, J. R. Nelson, R. J. Needs, Y. Li, H. Liu, Y. Zhang, Y. Ma, and F. Mauri, Phys. Rev. Lett. **114**, 157004 (2015).

[10] Ion Errea, Matteo Calandra, Chris J. Pickard, Joseph Nelson, Richard J. Needs, Yinwei Li, Hanyu Liu, Yunwei Zhang, Yanming Ma, Francesco Mauri. *Nature* 532, 81-84 (2016)

[11] L. P. Gor'kov and V. Z. Kresin,Pressure and high-Tc superconductivity in sulfur hydrides. *Scientific Reports* **6**, 25608 (2016), ISSN 2045-2322, URL http://dx.doi.org/10.1038/srep25608.

[12] A. Bianconi and T. Jarlborg,Superconductivity above the lowest Earth temperature in pressurized sulfur hydride, *EPL (Europhysics Letters)* **112**, 37001 (2015), URL http://dx.doi.org/10.1209/0295-5075/112/37001.

[13] T. Jarlborg and A. Bianconi, Breakdown of the migdal approximation at lifshitz transitions with giant zero-point motion in the H3S superconductor. *Scientific Reports* **6**, 24816 (2016), URL http://dx.doi.org/10.1038/srep24816.

[14] M. Lüders, M.A.L. Marques, N.N. Lathiotakis, A. Floris, G. Profeta, L. Fast, A. Continenza, S. Massidda, E.K.U. Gross. Ab-initio theory of superconductivity - I: Density functional formalism and approximate functionals. *Phys. Rev.* B72, 024545, 2005.

[15] A. José Flores-Livas, Antonio Sanna, E.K.U. Gross, arXiv:1501.06336, High temperature superconductivity in sulfur and selenium hydrides at high pressure.

[16] G. Gladstone, M.A.Jenson, J.R. Schrieffer. Superconductivity. Ed. By R.D. Parks. N.-Y., 1969.

[17] P.B.Allen, R.C. Dynes, *Phys.Rev.* B12, 905 (1975).

[18] W.L. McMillan, *Phys.Rev.*167**,** 331 (1968).

[19] E.J. Nicol, J.P. Carbotte. Comparison of pressurized sulfur hydride with conventional superconductors . *Phys. Rev.* **B 91**, 220507(R) (2015).

[20] Wataru Sano, Takashi Koretsune, Terumasa Tadano, Ryosuke Akashi, RyotaroArita. Effect of van Hove singularities on high-Tc superconductivity in H3S. *Phys. Rev.* **B 93**, 094525 (2016).

[21] G.M. Eliashberg, Interactions between electrons and lattice vibrations in a superconductor, *Sov. Phys. JETP* 11(3), 696-702 (1960)].

[22] G.M. Eliashberg, Temperature Greens function for electrons in a superconductor, *Sov. Phys. JETP* 12(5), 1000-1002 (1961).

[23] F.Marsiglio, J.P.Carbotte. Electron-phonon Superconductivity. In Superconductivity,Volume 1: Conventional and Unconventional Superconductors, K.H.Bennemann, Ketterson J.B. eds., 2008, 1568 pp. p.73-162.Springer, Berlin - Heidelberg.

[24] S. Engelsberg, J.R. Schriffer Coupled Electron-Phonon System, *Phys. Rev.* 1963. Vol. 131. № 3. P.993-1008.

[25] Scalapino D.G. //Superconductivity/ Ed. By R.D. Parks/ Dekker N-Y.1969. Vol. 1.

[26] Schrieffer J.R. Theory of Superconductivity (New York: W. Велjamin Inc., 1964).

[27] Morel A., Anderson P.W. Calculation of the Superconducting State Parameters with Retarded Electron-Phonon Interaction, *Phys. Rev.*, 125. № 4. P. 1263-1271(1962).

[28] S.V. Vonsovskii, Iu. A. Iziumov , E. Z. Kurmaev, Superconductivity of transition metals: their alloys and compounds. Springer-Verlag, 1982, 512 p.

[29] V.N. Grebenev, E.A. Mazur, Superconducting transition temperature of the nonideal A-15 structure. *Fizika nizkih temperatur,* 13, 478-483 (1987).

[30] Allen P.B., Mitrovic B.// Solid State Physics /Ed. By F. Seitz e.a. N.-Y.: Academic, 1984.

[31] D.J. Scalapino, J.R. Schrieffer, J.W. Wilkins, *Phys.Rev.* 148**,** 263 (1966).

[32] A.E. Karakozov, E.G. Maksimov, S.A. Mashkov. Effect of the frequency dependence of the electron-phonon interaction spectral function on the thermodynamic properties of superconductors . *JETP*, **41**, No. 5, p. 971 (1975).

[33] High Temperature superconductivity (New York; Consultants Bureau,1982), V.L. Ginzburg, D.A. Kirzhnits, Eds.

[34] F. Marsiglio, *Journ. of Low Temp. Phys.***87**, 659 (1992).

[35] W.E. Pickett, *Phys.Rev.* **B26**, 1186 (1982).

[36] A. S. Aleksandrov , V. N. Grebenev , E. A. Mazur , Violation of the Migdal theorem for a strongly coupled electron-phonon system, *Pis'ma v ZheTF,* 45, issue 7, p.357 (1987). Engl.transl. *JETP Lett.* **45** 455 (1987).

[37] T.Holstein, *Ann. Phys*. **29**, № 3. P. 410-535 (1964).

[38] G.Baym, *Ann.Phys.* **14**, 1, (1961).

[39] N.S. Gillis, *Phys.Rev.* **B1**, 1872, (1970).

[40] E.G. Maksimov. Effect of the frequency dependence of the electron-phonon interaction spectral function on the thermodynamic properties of superconductors . *Zh. Eksp. Teor. Fiz.,* **69,** 2236-2248 (1975); *Sov. Phys. JETP* 42, 1138-1143, (1976).

[41] A. B. Migdal Interaction between Electrons and Lattice Vibrations in a Normal Metal. *Sov. Phys. JETP*, **34** (7), 996-1001 (1958).

[42] L.D. Landay, E.M. Lifshitz, Statistical Physics, Part 2 (Pergamon, Oxford, 1980).

[43] E.A. Mazur, Phonon origin of low energy and high energy kinks in high temperature cuprate superconductors. *Europhysics Letters (EPL)*, **90**, 47005-47016 (2010).

[44] N. N. Degtyarenko, E. A. Mazur. Optimization of the superconducting phase of hydrogen sulfide. *Zh. Eksp. Teor. Fiz.,* **121**, Issue 6, 1067-1075 (2015).

[45] N.A Kudryashov, A.A. Kutukov, E. A. Mazur.The conductivity band reconstruction of metal hydrogen sulfide . *Zh. Eksp. Teor. Fiz,* **123,** 558 (2016).